\documentclass[12pt]{article}
\usepackage{enumerate}
\usepackage{amsfonts}
\usepackage{amsmath}
\usepackage{amssymb}
\usepackage{amsthm}
\usepackage{latexsym}
\usepackage{color}

\def\H{\mathcal{H}}

\def\S{\mathfrak{S}}
\def\C{\mathfrak{C}}
\def\T{\mathfrak{T}}

\def\B{\mathfrak{B}}

\newcommand{\supp}{\mathrm{supp}}
\newcommand{\rank}{\mathrm{rank}}
\newcommand{\id}{\mathrm{Id}}
\newcommand{\Tr}{\mathrm{Tr}}

\newcommand{\shs}{\hspace{1pt}}
\newcounter{defin}  \newcounter{lemma}  \newcounter{theorem}
\newcounter{property} \newcounter{corol}  \newcounter{remark} \newcounter{example}
\newenvironment{lemma}{\par\refstepcounter{lemma}     \textbf{Lemma \thelemma.} }{\rm\par}
\newenvironment{theorem}{\par\refstepcounter{theorem}     \textbf{Theorem \thetheorem.}\ }{\rm\par}
\newenvironment{property}{\par\refstepcounter{property}     \textbf{Proposition \theproperty.}\ }{\rm\par}
\newenvironment{corollary}{\par\refstepcounter{corol}     \textbf{Corollary \thecorol.} }{\rm\par}

\newenvironment{remark}{\par\refstepcounter{remark}     \textbf{Remark \theremark.}}{\rm\par}
\newenvironment{example}{\par\refstepcounter{example}     \textbf{Example \theexample.}}{\rm\par}

\begin{document}

\title{Approximation of multipartite quantum states and the relative entropy of entanglement}
\author{M.E.~Shirokov\footnote{Steklov Mathematical Institute, Moscow, Russia, email:msh@mi.ras.ru}}
\date{}
\maketitle
\begin{abstract}
Special approximation technique for analysis of different characteristics of states of multipartite
infinite-dimensional  quantum systems is proposed and applied to study of the relative entropy of entanglement and its regularisation.

We prove several results about analytical properties of the multipartite relative entropy of entanglement and its regularization (the lower semicontinuity on  wide class of states, the uniform continuity under the energy constraints, etc.).

We establish a finite-dimensional approximation property for the relative entropy of entanglement and its regularization
that allows to generalize to the infinite-dimensional case the results proved in the finite-dimensional settings.
\end{abstract}

\tableofcontents

\section{Introduction}

A specific feature of infinite-dimensional quantum systems is singular properties (discontinuity, infinite values, etc.) of basic characteristics
of quantum states. So, for strict mathematical analysis of infinite-dimensional quantum systems it is necessary to apply special techniques, in particular, approximation
techniques to overcome the problems arising from these singularities.

In this article we develop the approximation technique proposed in \cite{FAP} in the one-party case to analysis of characteristics of multipartite infinite-dimensional quantum systems.
A central notion of this technique is the finite-dimensional approximation property of a state (briefly called the FA-property). According to \cite{FAP}
a quantum state $\rho$ with the spectrum $\{\lambda_i\}$ has the FA-property if there exists of a sequence $\{g_i\}$ of nonnegative numbers such that
\begin{equation}\label{FA-d}
\sum_{i=1}^{+\infty}\lambda_i g_i<+\infty\quad \textrm{and} \quad \lim_{\beta\to 0^+}\left[\sum_{i=1}^{+\infty}e^{-\beta g_i}\right]^{\beta}=1.
\end{equation}

It is shown in \cite{FAP} that the FA-property of a state $\rho$ implies the finiteness of its entropy, but the converse
implication remains an open question. There is a simple sufficient condition for the FA-property: it holds for a state $\rho$ with the spectrum $\{\lambda_i\}$ provided that
\begin{equation*}
\sum_{i=1}^{+\infty}\lambda_i \ln^q i<+\infty\quad \textrm{for some} \quad q>2.
\end{equation*}
This condition shows that the FA-property holds for all states whose eigenvalues tend to zero faster than $[i\ln^q i]^{-1}$ as $i\to+\infty$ for some $q>3$, in particular, it holds
for all Gaussian states playing essential role in quantum information theory.

Many characteristics of a state of a $n$-partite quantum system $A_1...A_n$ have the form of a function
$$
f(\rho\,|\,p_1,...,p_l)
$$
on the set $\S(\H_{A_1...A_n})$ of states of this system depending on some parameters  $p_1,...,p_l$ (other states, quantum channels, quantum measurements, etc.).
If $\rho$ is a state of infinite-dimensional $n$-partite quantum system $A_1...A_n$ then we may approximate it by the sequence of states
\begin{equation}\label{rho-r}
\rho_r=Q_r\rho\shs Q_r [\Tr Q_r\rho\shs]^{-1},\quad  Q_r=P_r^1\otimes...\otimes P_r^n,
\end{equation}
where $P_r^s$ is the spectral projector of $\rho_{A_s}$ corresponding to its $r$ maximal eigenvalues  (taking the multiplicity into account).
Naturally, the question arises under what conditions
$$
f(\rho_r\,|\,p_1,...,p_l)\quad \textrm{tends to}\quad f(\rho\,|\,p_1,...,p_l)\quad \textrm{as} \;\; r\to\infty
$$
uniformly on $p_1,...,p_l$.  Our main technical result asserts that for given $m\leq n$  the FA-property of the marginal
states $\rho_{A_1}$,..., $\rho_{A_m}$ of a  state
$\rho$ in $\S(\H_{A_1...A_n})$ guarantees this uniform convergence for a wide class of functions $f$ (depending on $m$).
In fact, the more general assertion is valid in which the states $\rho_r$ are defined by formula (\ref{rho-r}) via the operators $Q_r=P_r^{s_1}\otimes...\otimes P_r^{s_l}\otimes I_{R}$, where  $\{s_1,...,s_l\}$ is any subset of $\{1,...,n\}$ and $R=A_1...A_n\setminus A_{s_1}...A_{s_l}$ (Theorem  \ref{main} in Section 3).

The above result is used in this article for analysis of the relative entropy of entanglement and its regularization in infinite-dimensional $n$-partite quantum systems
--  the basic entanglement measures used in quantum information theory \cite{E_R-D,V&P,4H}. Mathematically, particular problems of studying the relative entropy of entanglement in infinite dimensions
are related to its definition involving the infimum of a lower semicontinuous function (the quantum relative entropy) over a noncompact set (the set of separable states). One of these
problems is a proof of lower semicontinuity of the relative entropy of entanglement, which is a desirable property of an entanglement measure in infinite-dimensional composite systems.

The proposed approximation technique allows to establish  the lower semicontinuity of the relative entropy of entanglement
$E_R$ and its regularization $E^{\infty}_R$ on the subset $\S_*(\H_{A_1...A_n})$ of $\S(\H_{A_1...A_n})$ consisting of states $\rho$ that have at least $n-1$
marginal states with the FA-property. Moreover, it is  proved that
\begin{equation*}
\liminf_{k\to+\infty} E^*_R(\rho_k)\geq E^*_R(\rho_0),\quad E_R^*=E_R,E^{\infty}_R,
\end{equation*}
for arbitrary sequence $\{\rho_k\}$ converging to a state $\rho_0$ in $\S_*(\H_{A_1...A_n})$.

The approximation technique is used also to obtain several observations concerning definition of the relative entropy of entanglement. In particular, we
prove that for all states in $\S_*(\H_{A_1...A_n})$  this quantity can be defined as the relative entropy distance to the set of all finitely decomposable separable states (despite the
existence of finitely-non-decomposable and countably-non-decomposable separable states in $\S(\H_{A_1...A_n})$).

We establish a finite-dimensional approximation property for the relative entropy of entanglement and its regularization
that allows to generalize to the infinite-dimensional case the results proved in the finite-dimensional settings. Examples of using this property are presented.

Finally, we consider energy-constrained versions of the $n$-partite relative entropy of entanglement. It is proved, in particular, that for any state in $\S(\H_{A_1...A_n})$ with finite energy the infimum in the definition of the relative entropy of entanglement can be taken over all finitely-decomposable separable states with finite energy (provided that the Hamiltonians of individual subsystems satisfy a particular condition).

\section{Preliminaries}

Let $\mathcal{H}$ be a separable Hilbert space,
$\mathfrak{B}(\mathcal{H})$ the algebra of all bounded operators on $\mathcal{H}$ with the operator norm $\|\cdot\|$ and $\mathfrak{T}( \mathcal{H})$ the
Banach space of all trace-class
operators on $\mathcal{H}$  with the trace norm $\|\!\cdot\!\|_1$. Let
$\mathfrak{S}(\mathcal{H})$ be  the set of quantum states (positive operators
in $\mathfrak{T}(\mathcal{H})$ with unit trace) \cite{H-SCI,N&Ch,Wilde}.

Denote by $I_{\mathcal{H}}$ the unit operator on a Hilbert space
$\mathcal{H}$ and by $\id_{\mathcal{\H}}$ the identity
transformation of the Banach space $\mathfrak{T}(\mathcal{H})$.\smallskip

Following \cite{FAP} we will say that a state $\rho$ in $\S(\H)$ with the spectrum $\{\lambda_i\}$ has the FA-property if there exists of a sequence $\{g_i\}$ of nonnegative numbers such that
(\ref{FA-d}) holds. We will denote by $\S_{\rm \textsf{FA}\!}(\H)$ the set of all states in $\S(\H)$ having the FA-property. It follows from Corollary 3 in \cite{FAP} that
$\S_{\rm \textsf{FA}\!}(\H)$ is a face of the convex set $\S(\H)$.\footnote{It means that the set $\S_{\rm \textsf{FA}\!}(\H)$ is convex and contains any segment from $\S(\H)$ provided that
it contains at least one internal point of this segment.}\smallskip

The \emph{von Neumann entropy} of a quantum state
$\rho \in \mathfrak{S}(\H)$ is  defined by the formula
$H(\rho)=\operatorname{Tr}\eta(\rho)$, where  $\eta(x)=-x\ln x$ for $x>0$
and $\eta(0)=0$. It is a concave lower semicontinuous function on the set~$\mathfrak{S}(\H)$ taking values in~$[0,+\infty]$ \cite{H-SCI,L-2,W}.
The von Neumann entropy satisfies the inequality
\begin{equation}\label{w-k-ineq}
H(p\rho+(1-p)\sigma)\leq pH(\rho)+(1-p)H(\sigma)+h_2(p)
\end{equation}
valid for any states  $\rho$ and $\sigma$ in $\S(\H)$ and $p\in(0,1)$, where $\,h_2(p)=\eta(p)+\eta(1-p)\,$ is the binary entropy \cite{N&Ch,Wilde}. Note that $H(\rho)$ is finite for any state $\rho$ in $\S_{\rm \textsf{FA}\!}(\H)$ \cite{FAP}.\smallskip

We will use the Lindblad extension of the \emph{quantum relative entropy}
defined for any  positive operators $\rho$ and
$\sigma$ in $\mathfrak{T}(\mathcal{H})$ as follows
\begin{equation}\label{qre-L}
H(\rho\shs\|\shs\sigma)=\sum_i\langle
i|\,\rho\ln\rho-\rho\ln\sigma\,|i\rangle+\Tr\sigma-\Tr\rho,
\end{equation}
where $\{|i\rangle\}$ is the orthonormal basis of
eigenvectors of the operator $\rho$ and it is assumed that
$H(\rho\shs\|\shs\sigma)=+\infty$ if $\,\mathrm{supp}\rho\shs$ is not
contained in $\shs\mathrm{supp}\shs\sigma$ \cite{L-2}.\footnote{The support $\mathrm{supp}\rho$ of a positive trace class operator $\rho$ is the closed subspace spanned by the eigenvectors of $\rho$ corresponding to its positive eigenvalues.}
\smallskip

The \emph{quantum conditional entropy}
\begin{equation*}
H(A|B)_{\rho}=H(\rho)-H(\rho_{\shs B})
\end{equation*}
of a  state $\rho$ of a
bipartite quantum system $AB$ with finite marginal entropies $H(\rho_A)$ and $H(\rho_B)$ is essentially used in analysis of quantum systems \cite{H-SCI,Wilde}. It
can be extended to the set of all states $\rho$ with finite $H(\rho_A)$ by the formula
\begin{equation}\label{ce-ext}
H(A|B)_{\rho}=H(\rho_{A})-H(\rho\shs\Vert\shs\rho_{A}\otimes
\rho_{B})
\end{equation}
proposed in \cite{Kuz}, where it is shown that this extension  possesses all basic properties of the quantum conditional entropy valid in finite dimensions.  \smallskip

The \emph{quantum mutual information} (QMI) of a state $\,\rho\,$ of a multipartite quantum system $A_1 \ldots
A_n$ is defined as (cf.\cite{L-mi,Herbut,NQD})
\begin{equation}\label{mi-mpd}
     I(A_1\!:\ldots:\!A_n)_{\rho}\doteq
    H(\rho\,\|\,\rho_{A_{1}}\otimes\cdots\otimes\rho_{A_{n}})=\sum_{s=1}^n H(\rho_{A_{s}})-H(\rho),
\end{equation}
where the second formula is valid if $H(\rho)<+\infty$. It is easy to show (cf.\cite{CBM}) that
\begin{equation}\label{nMI-UB}
  I(A_1\!:\ldots:\!A_n)_{\rho}\leq 2\sum_{s=1}^{n-1}H(\rho_{A_s}).
\end{equation}
Similar upper bound holds with any other $n-1$ marginal entropies of the state $\rho$. \smallskip

For any positive (semi-definite) densely defined operator{\footnote{We assume that a positive operator is a self-adjoint operator \cite{R&S}.}} $G$ on a Hilbert space $\H$ and any positive operator $\rho$ in $\T(\H)$ we assume that
\begin{equation}\label{Tr-def}
\Tr G \rho=\sup_n\mathrm{Tr} P_n G\rho\leq+\infty,
\end{equation}
where $P_n$ is the spectral projector of $G$ corresponding to the interval $[0,n]$. Then
\begin{equation}\label{C-set}
\mathfrak{C}_{G,E}=\left\{ \rho\in\mathfrak{S}(\H)\,|\,\mathrm{Tr} G\rho\leq E \right\}
\end{equation}
is a closed convex nonempty subset of $\mathfrak{S}(\H)$ for any $E$ exceeding the infimum of the spectrum of $G$. If
$G$ is treated as  Hamiltonian of the quantum system associated with the space $\H$ then
$\mathfrak{C}_{G,E}$  is the set of states with the mean energy not exceeding $E$.\smallskip

We will pay a special attention to the class of  unbounded densely defined positive operators on $\H$ having discrete spectrum of finite multiplicity.
In Dirac's notations any such operator $G$ can be represented as follows
\begin{equation}\label{H-rep}
G=\sum_{i=1}^{+\infty} g_i|\tau_i\rangle\langle \tau_i|
\end{equation}
on the domain $\mathcal{D}(G)=\{ \varphi\in\H\,|\,\sum_{i=1}^{+\infty} g^2_i|\langle\tau_i|\varphi\rangle|^2<+\infty\}$, where
$\left\{\tau_i\right\}_{i=1}^{+\infty}$ is the orthonormal basis of eigenvectors of $G$
corresponding to the nondecreasing sequence $\left\{\smash{g_i}\right\}_{i=1}^{+\infty}$ of eigenvalues
tending to $+\infty$.

It is well known that the von Neumann entropy is continuous on the set $\mathfrak{C}_{G,E}$ for any $E$ if (and only if) the operator  $G$ satisfies  the condition
\begin{equation}\label{H-cond}
  \mathrm{Tr}\, e^{-\beta G}<+\infty\quad\textrm{for all}\;\beta>0
\end{equation}
and that the maximal value of the entropy on this set is achieved at the \emph{Gibbs state} $\gamma(E)\doteq e^{-\beta(E) G}/\mathrm{Tr} e^{-\beta(E) G}$, where the parameter $\beta(E)$ is determined by the equality $\mathrm{Tr} G e^{-\beta(E) G}=E\mathrm{Tr} e^{-\beta(E) G}$ \cite{W}. Condition (\ref{H-cond}) implies that $G$ is an unbounded operator having  discrete spectrum of finite multiplicity, i.e. it has form (\ref{H-rep}). So, by the Lemma in \cite{H-c-w-c} the set $\mathfrak{C}_{G,E}$ defined in (\ref{C-set}) is compact for any $E$.

We will often consider operators $G$ satisfying the condition
\begin{equation}\label{H-cond+}
  \lim_{\beta\rightarrow0^+}\left[\mathrm{Tr}\, e^{-\beta G}\right]^{\beta}=1,
\end{equation}
which is slightly stronger than condition (\ref{H-cond}). In terms of the sequence $\{g_i\}$ of eigenvalues of $G$
condition (\ref{H-cond}) means that $\lim_{i\rightarrow\infty}g_i/\ln i=+\infty$, while condition (\ref{H-cond+}) is valid  if $\;\liminf_{i\rightarrow\infty} g_i/\ln^q i>0\,$ for some $\,q>2$ \cite[Proposition 1]{AFM}. By Lemma 1 in \cite{AFM} condition (\ref{H-cond+}) holds if and only if
\begin{equation}\label{H-cond++}
  F_{G}(E)\doteq\sup_{\rho\in\mathfrak{C}_{G,E}}H(\rho)=o\shs(\sqrt{E})\quad\textrm{as}\quad E\rightarrow+\infty.
\end{equation}
It is essential that condition (\ref{H-cond+}) is valid for the Hamiltonians of many real quantum systems \cite{Datta,AFM}.

\section{Approximation of multipartite quantum states}

In this section we will obtain  the $n$-partite version of Theorem 2 in \cite{FAP}. It turns out that many important characteristics of a state of a $n$-partite quantum systems  belong to one of the classes $\widehat{L}^{m}_n(C,D)$ and $N^{m}_{n,s}(C,D)$ introduced in \cite{CBM} and described below.

\smallskip

Let $m\leq n$ and $L^{m}_n(C,D)$ be the class of all functions $f$ on the set\footnote{We assume that $A_1$,...,$A_n$ are infinite-dimensional quantum systems.}
$$
\S_m(\H_{A_1..A_n})\doteq\left\{\rho\in\S(\H_{A_1..A_n})\,|\,H(\rho_{A_1}),..., H(\rho_{A_m})<+\infty\shs\right\}
$$
such that
\begin{equation}\label{F-p-1}
-a_f h_2(p)\leq f(p\rho+(1-p)\sigma)-p f(\rho)-(1-p)f(\sigma)\leq b_fh_2(p)
\end{equation}
for any states $\rho$  and $\sigma$ in $\S_m(\H_{A_1..A_n})$ and any $p\in(0,1)$ and
\begin{equation}\label{F-p-2}
-c^-_f S_m(\rho)\leq f(\rho)\leq c^+_f S_m(\rho),\quad \quad S_m(\rho)=\sum_{s=1}^m H(\rho_{A_s}),
\end{equation}
for any state $\rho$ in $\S_m(\H_{A_1..A_n})$, where $h_2$ is the binary entropy (defined after (\ref{w-k-ineq})),
$a_f,b_f$ and $c^\pm_f$ are nonnegative numbers such that $a_f+b_f=D$ and $c^-_f+c^+_f=C$.

Let $\widehat{L}^{m}_n(C,D)$ be the class containing all functions in $L^{m}_n(C,D)$
and all the functions of the form
$$
f(\rho)=\sup_{\lambda}f_{\lambda}(\rho)\quad \textrm{and} \quad f(\rho)=\inf_{\lambda}f_{\lambda}(\rho),
$$
where $\{f_{\lambda}\}$ is a family of functions in $L^{m}_n(C,D)$.

Let $N^{m}_{n,s}(C,D)$ be the class of all functions $f$ on the set $\S_m(\H_{A_{1}..A_{n}})$
defined by the expression
\begin{equation*}
f(\rho)\doteq \inf_{\hat{\rho}\in \mathfrak{M}_s(\rho)}h(\hat{\rho})
\end{equation*}
via particular function $h$ in $\widehat{L}^{m}_{n+l}(C,D)$ for some $l>0$, where:
\begin{itemize}
  \item $\mathfrak{M}_1(\rho)$ is the set of all extensions of $\rho$ to a state of $A_{1}...A_{n+l}$;
  \item $\mathfrak{M}_2(\rho)$ is the set of all extensions of $\rho$ having the form
  \begin{equation}\label{c-ext}
       \hat{\rho}=\sum_i p_i \rho_i\otimes |i\rangle\langle i|,
  \end{equation}
  where $\{\rho_i\}$ is a collection of states in $\S(\H_{A_{1}..A_{n}})$, $\{p_i\}$ is a probability distribution and $\{|i\rangle\}$ is an orthonormal basis in $\H_{A_{n+1}}$ (in this case $l=1$);
  \item $\mathfrak{M}_3(\rho)$ is the set of all extensions of $\rho$ having the form (\ref{c-ext}) in which $\{\rho_i\}$ is a collection of \emph{pure} states in $\S(\H_{A_{1}..A_{n}})$.
\end{itemize}

A noncomplete list of characteristics belonging to one of the classes $\widehat{L}^{m}_n(C,D)$ and $N^{m}_{n,s}(C,D)$ includes the von Neumann entropy, the conditional entropy, the $n$-partite quantum (conditional) mutual information, the one way classical correlation, the quantum discord, the mutual and coherent informations of a quantum channel,  the
information gain of a quantum measurement with and without quantum side information, the $n$-partite relative entropy of entanglement, the quantum topological entropy and its\break  $n$-partite generalization, the bipartite entanglement of formation, the $n$-partite squashed entanglement and c-squashed entanglement, the conditional entanglement of
mutual information and other conditional entanglement measures obtained via some function from one of the classes $\widehat{L}^{m}_{n+l}(C,D)$, $l>0$  \cite[Section 3]{CBM}.\smallskip

To formulate our main  technical result introduce the approximation map $\Lambda^{s_1,...,s_l}_r$ on the space
$\S(\H_{A_1...A_n})$ determined by a set $\{s_1,...,s_l\}\subseteq \{1,...,n\}$ as follows
\begin{equation}\label{ap-map}
\Lambda^{s_1,...,s_l}_r(\rho)=Q_r\rho\shs Q_r [\Tr Q_r\rho\shs]^{-1},\quad  Q_r=P_r^{s_1}\otimes...\otimes P_r^{s_l}\otimes I_{R},
\end{equation}
where $P_r^s$ is the spectral projector of $\rho_{A_s}$ corresponding to its $r$ maximal eigenvalues  (taking the multiplicity into account) and
$R=A_1...A_n\setminus A_{s_1}...A_{s_l}$. \smallskip\pagebreak

\begin{theorem}\label{main} \emph{Let $\rho$ be a state in $\S(\H_{A_1..A_n})$ such that $\rho_{A_s}\in\S_{\!\rm \textsf{FA}\!}(\H_{A_s})$ for $s=\overline{1,m}$, $m\leq n$.\footnote{$\S_{\rm \textsf{FA}\!}(\H)$ is the set of all states in $\S(\H)$ having the FA-property (see Section 2).}
Let $\{s_1,...,s_l\}$ be a given subset of $\{1,...,n\}$. Then for given nonnegative numbers $C$ and $D$ there
is a natural number $r_0$ and vanishing sequences $\{Y^L_{C,D}(r)\}_{r\in\mathbb{N}}$ and $\{Y^N_{C,D}(r)\}_{r\in\mathbb{N}}$ such that
$$
|f(\Lambda^{s_1,...,s_l}_r(\rho))-f(\rho)|\leq Y^L_{C,D}(r)\quad \textit{and} \quad|f'(\Lambda^{s_1,...,s_l}_r(\rho))-f'(\rho)|\leq Y^N_{C,D}(r) \quad \forall r\geq r_0
$$
for any function $f$ in $\widehat{L}^{m}_n(C,D)$ and
any function $f'$ in $N^{m}_{n,s}(C,D)$.}

\emph{The number $r_0$ and the sequences $\{Y^L_{C,D}(r)\}_{r\in\mathbb{N}}$ and $\{Y^N_{C,D}(r)\}_{r\in\mathbb{N}}$ are completely determined by the states $\rho_{A_s}$, $s\in \{s_1,...,s_l\}\cup\{1,...,m\}$, and do not depend on $n$.}
\end{theorem}\medskip

\emph{Proof.} For each natural $s$ in $[1,m]$ let $\{g^s_i\}_i$ be a sequence of nonnegative numbers such that
\begin{equation}\label{FA++}
\sum_{i=1}^{+\infty}\lambda^s_i g^s_i<+\infty\quad \textrm{and} \quad \lim_{\beta\to 0^+}\left[\sum_{i=1}^{+\infty}e^{-\beta g^s_i}\right]^{\beta}=1,
\end{equation}
where $\{\lambda^s_i\}$ is the spectrum of the state $\rho_{A_s}$ taking in the non-increasing order. We may
assume that all the sequences  $\{g^s_i\}_i$ are non-decreasing and that $g_1^s=0$ for all $s$. For each $s$ consider the positive operator
$$
G_{\!A_s}=\sum_{i=1}^{+\infty}g^s_i|\varphi^s_i\rangle\langle \varphi^s_i|
$$
on $\H_{A_s}$, where $\{\varphi^s_i\}$ is the basis of eigenvectors of $\rho_{A_s}$ corresponding to the sequence $\{\lambda^s_i\}$ of eigenvalues. Then
the operator $G_{\!A_s}$ satisfies condition (\ref{H-cond+}) and
$$
E_s\doteq \Tr G_{\!A_s}\rho_{A_s}=\sum_{i=1}^{+\infty}\lambda^s_i g^s_i<+\infty.
$$

For each natural $s$ in $[1,n]$ let $P_r^s=\sum_{i=1}^{r}|\varphi^s_i\rangle\langle \varphi^s_i|$ be the spectral projector of $\rho_{A_s}$ corresponding to its $r$ maximal eigenvalues  (taking the multiplicity into account)
and $\bar{P}_r^s=I_{A_s}-P_r^s$. Then
\begin{equation}\label{p-ineq}
\Tr Q_r\rho\geq 1-\sum_{j=1}^l\Tr\bar{P}_r^{s_j}\rho_{A_{s_j}},
\end{equation}
where $Q_r$ is the projector defined in (\ref{ap-map}).  To prove this inequality  note that
$$
\left|\Tr Q_r^{j-1}\rho - \Tr Q_r^{j}\rho\shs\right|
\leq \|Q_r^{j-1}\|\Tr[\bar{P}_r^{s_j}\otimes I_{A_1...A_n\setminus A_{s_j}}]\rho
=\Tr\bar{P}_r^{s_j}\rho_{A_{s_j}},\quad j=\overline{1,l},
$$
where $Q_r^0=I_{A_1..A_n}$, $\,Q_r^j=P_r^{s_1}\otimes...\otimes P_r^{s_j}\otimes I_{R_j}$,  $R_j=A_1...A_n\setminus A_{s_1}...A_{s_j}$. Hence
$$
1-\Tr Q_r\rho\leq \sum_{j=1}^l \left|\Tr Q_r^{j-1}\rho - \Tr Q_r^{j}\rho\shs\right|\leq \sum_{j=1}^l\Tr\bar{P}_r^{s_j}\rho_{A_{s_j}}.
$$

By the construction of the state $\Lambda^{s_1,...,s_l}_r(\rho)$ we have
$$
c_r[\Lambda^{s_1,...,s_l}_r(\rho)]_{A_s}\leq \rho_{A_s},\quad c_r=\Tr Q_r\rho,
$$
for any $s$. Hence, it follows from (\ref{p-ineq}) that
\begin{equation}\label{e-cond}
  \sum_{s=1}^{m}\Tr G_{A_s}[\Lambda^{s_1,...,s_l}_r(\rho)]_{A_s}\leq c^{-1}_r\sum_{s=1}^{m}\Tr G_{A_s}\rho_{A_s}=c^{-1}_r E_S\leq \frac{E_S}{1-\sum_{j=1}^l\Tr\bar{P}_r^{s_j}\rho_{A_{s_j}}},
\end{equation}
where $E_{S}=E_1+...+E_m$.

By using Winter's gentle measurement lemma (cf.\cite{Wilde}) we obtain
\begin{equation}\label{n-est}
\|\rho-\Lambda^{s_1,...,s_l}_r(\rho)\|_1\leq2\sqrt{\Tr \bar{Q}_r\rho}\leq 2\sqrt{\sum_{j=1}^l\Tr\bar{P}_r^{s_j}\rho_{A_{s_j}}},
\end{equation}
where $\,\bar{Q}_r=I_{A_1...A_n}-Q_r\,$ and the last inequality follows from (\ref{p-ineq}).

Let $A^m\doteq A_1...A_m$. Since all the operators $G_{A_1}$,...,$G_{A_m}$ satisfy condition  (\ref{H-cond+}), the operator
\begin{equation*}
G_{A^m}=G_{A_1}\otimes I_{A_2}\otimes...\otimes I_{A_m}+\cdots+I_{A_1}\otimes... \otimes I_{A_{m-1}}\otimes G_{A_m}.
\end{equation*}
satisfies the same condition
by Lemma 2 in \cite{CBM}. By Lemma 1 in \cite{AFM} this implies that
\begin{equation}\label{F-fun}
F_{G_{A^m}}(E)\doteq\sup_{\Tr G_{A^m}\sigma\leq E}H(\sigma)=o(\sqrt{E})\quad \textrm{as} \;\; E\to+\infty,
\end{equation}
where the supremum is over all states $\sigma$ in $\S(\H_{A^m})$ such that $\Tr G_{A^m}\sigma\leq E$.

Let $r_0$ be the minimal positive integer  such that $\delta_r\doteq \sqrt{\sum_{j=1}^l\Tr\bar{P}_r^{s_j}\rho_{A_{s_j}}}\leq 1/2$ for all $r\geq r_0$.
The equality $\,\sum_{s=1}^{m}\Tr G_{A_s}\rho_{A_s}=E_S\,$ and inequality (\ref{e-cond}) allow to apply  Theorem 1 in \cite{CBM}
to the states $\rho$ and $\Lambda^{s_1,...,s_l}_r(\rho)$ for all $r\geq r_0$. By this theorem it follows from (\ref{n-est}) that
\begin{equation}\label{d-one}
  |f(\rho)-f(\Lambda^{s_1,...,s_l}_r(\rho))|\leq C\sqrt{2\delta_r}F_{G_{A^m}}\!\!\left[\frac{4E_S}{3\delta_r}\right]+Dg(\sqrt{2\delta_r}) \qquad \forall r\geq r_0
\end{equation}
for any function $f\in \widehat{L}^{m}_n(C,D)$ and
\begin{equation}\label{d-one+}
  |f'(\rho)-f'(\Lambda^{s_1,...,s_l}_r(\rho))|\leq C\sqrt{2\delta'_r}F_{G_{A^m}}\!\!\left[\frac{4E_S}{3\delta'_r}\right]+Dg(\sqrt{2\delta'_r}) \qquad \forall r\geq r_0,
\end{equation}
where $\delta'_r=\sqrt{\delta_r(2-\delta_r)}$, for any function $f'\in N^{m}_{n,s}(C,D)$.

Since $\delta_r$ tends to zero as $\,r\to+\infty$, by denoting the right hand sides  of (\ref{d-one}) and (\ref{d-one+}) by $Y^L_{CD}(r)$ and $Y^N_{CD}(r)$ correspondingly and taking (\ref{F-fun}) into account we obtain the main assertion of the theorem.\smallskip

The last assertion follows from the above proof. $\square$ \smallskip

\begin{example}\label{m-ex} Let $f_{\Phi_1...\Phi_n}(\rho)=I(B_1\!:...:\!B_n)_{\Phi_1\otimes...\otimes\Phi_n(\rho)}$, where $\Phi_1:A_1\rightarrow B_1$,..., $\Phi_n:A_n\rightarrow B_n$
are arbitrary quantum channels. Inequality (10) in \cite{CBM} (with trivial system $C$), upper bound (\ref{nMI-UB}), the nonnegativity of the quantum mutual information and its monotonicity  under local channels imply that the function $f_{\Phi_1...\Phi_n}$ belongs to the class $L_{n}^{n-1}(2,n)$ for any channels $\Phi_1,...,\Phi_n$.\footnote{The obvious inequality $\,I(A_1\!:\ldots:\!A_n)_{\rho}\leq \sum_{s=1}^{n}H(\rho_{A_s})\,$ implies that this function also belongs to the class $L_{n}^{n}(1,n)$.} Thus, if the marginal states $\rho_{A_1}$,...,$\rho_{A_{n-1}}$ have the FA-property then, by Theorem
\ref{main}, for any subset $\{s_1,...,s_l\}$ of $\{1,...,n\}$ there exist a natural number $r_0$ and a vanishing sequence $\{Y(r)\}$  such that
\begin{equation}\label{mi-uc}
|I(B_1\!:...:\!B_n)_{\Phi_1\otimes...\otimes\Phi_n(\rho_r)}-I(B_1\!:...:\!B_n)_{\Phi_1\otimes...\otimes\Phi_n(\rho)}|\leq Y(r)  \qquad \forall r\geq r_0,
\end{equation}
where $\rho_r=\Lambda^{s_1,...,s_l}_r(\rho)$, for any channels $\Phi_1,...,\Phi_n$. \smallskip

The last assertion of Theorem \ref{main} shows that inequality (\ref{mi-uc}) remains valid with
$\rho$ and $\rho_r=\Lambda^{s_1,...,s_l}_r(\rho)$ replaced by $\sigma$ and $\sigma_r=\Lambda^{s_1,...,s_l}_r(\sigma)$, where
$\sigma$ is any state in $\S(\H_{A_1..A_n})$ such that $\sigma_{A_k}=\rho_{A_k}$ for each $k\in\{s_1,...,s_l\}\cup\{1,...,n-1\}$.  \smallskip

By using the above observation with $\{s_1,...,s_l\}=\{1,...,n\}$ it is easy to prove continuity of the function
$$
(\Phi_1,...,\Phi_n)\mapsto I(B_1\!:...:\!B_n)_{\Phi_1\otimes...\otimes\Phi_n(\rho)}
$$
on the set of all tuples $(\Phi_1,...,\Phi_n)$ of channels equipped with the topology of pairwise strong convergence provided that
the FA-property holds for at least $n-1$ marginal states of the state $\rho$. By noting that the FA-property implies finiteness of the entropy, this assertion can be also obtained from Proposition 3 in \cite{LSE} proved in a completely different way.
\end{example}

\smallskip

\smallskip

\section{The relative entropy of entanglement and its regularization in infinite dimensions}

The relative entropy of entanglement is one of the main entanglement measures in finite-dimensional multipartite quantum systems. For a state $\rho$ of a system $A_1...A_n$ it is defined as
\begin{equation}\label{ree-def}
  E_R(\rho)=\inf_{\sigma\in\S_{\mathrm{s}}(\H_{A_1...A_n})}H(\rho\shs\|\shs\sigma),
\end{equation}
where $\S_{\mathrm{s}}(\H_{A_1...A_n})$ is the set of separable\footnote{Here and in what follows speaking about separable states we mean \emph{full} separable states \cite{4H}.} (nonentangled) states in $\S(\H_{A_1...A_n})$ defined as the convex hull of all product states $\rho_1\otimes \cdots\otimes\rho_n$, $\rho_s\in\S(\H_{A_s})$, $s=\overline{1,n}$ \cite{E_R-D,V&P,4H}.\footnote{A detailed description of the works devoted to the relative entropy of entanglement and related entanglement measures in bipartite quantum systems is given in Section 5.8 in \cite{Wilde-new}. I would be grateful for any references concerning the relative entropy of entanglement in  multipartite quantum systems.}

The relative entropy of entanglement possesses basic properties of entanglement measures (convexity, LOCC-monotonicity, asymptotic continuity, etc.) and satisfies
the  inequality
\begin{equation}\label{RE-LAA}
p E_R(\rho)+(1-p)E_R(\sigma)\leq E_R(p\rho+(1-p)\sigma)+h_2(p),
\end{equation}
valid for any states $\rho$  and $\sigma$ in $\S(\H_{A_1...A_n})$ and any $p\in[0,1]$, where $h_2$ is the binary entropy \cite{W-CB}.
In \cite{ERUB} it is proved that
\begin{equation}\label{ER-UB}
E_R(\rho)\leq  \sum_{s=1}^{n-1}H(\rho_{A_{s}}).
\end{equation}
Since this upper bound holds with arbitrary $n-1$ subsystems of $A_1...A_n$ (instead of $A_1,...,A_{n-1}$), it is easy to show that
\begin{equation}\label{ER-UB+}
  E_R(\rho)\leq \,\frac{n-1}{n}\sum_{s=1}^{n}H(\rho_{A_s}).
\end{equation}

The relative entropy of entanglement is nonadditive. Its regularization is defined in the standard way:
\begin{equation}\label{ree-r-def}
  E^{\infty}_R(\rho)=\lim_{m\rightarrow+\infty}m^{-1}E_R(\rho^{\otimes m}),
\end{equation}
where $\rho^{\otimes m}$ is treated as a state of the $n$-partite quantum system $A^m_1...A^m_n$.\smallskip

Below we consider two ways to generalize the finite-dimensional relative entropy of entanglement
to states of an infinite-dimensional $n$-partite quantum system. Then, in Section 4.3, we analyse  relations
between these generalizations and consider their analytical properties. We also explore analytical properties of the
regularized relative entropy of entanglement in the infinite-dimensional settings.

\subsection{Direct definition of $E_R$}

Definition (\ref{ree-def}) is valid in the case of infinite-dimensional quantum system $A_1...A_n$. One should only to note that in this case the set $\S_{\mathrm{s}}(\H_{A_1...A_n})$ is defined as the convex closure of all product states in $\S(\H_{A_1...A_n})$. It is essential that  inequality
(\ref{RE-LAA}) and upper bounds (\ref{ER-UB}) and (\ref{ER-UB+}) remain valid in this case provided that all the involved quantities are finite.
Inequality (\ref{RE-LAA})  in the infinite-dimensional settings is proved in \cite[Lemma 6]{AFM}. Upper bounds (\ref{ER-UB}) and (\ref{ER-UB+}) can be easily obtained from Lemma \ref{Omega} in the Appendix by using the $\sigma$-convexity of $E_R$ (which directly follows from the joint convexity and lower semicontinuity of the quantum relative entropy, see Theorem \ref{RE-LS}E below).

Inequalities (\ref{RE-LAA}), (\ref{ER-UB}) and (\ref{ER-UB+}) along with the convexity of $E_R$ show that
the function $E_R$ belongs to the classes  $L_n^{n-1}(1,1)$ and  $L_n^{n}(C_n,1)$, where $C_n=(n-1)/n$.\footnote{These classes are described in Section 3.}

Similar to other entanglement measures the relative entropy of entanglement is not continuous on the whole set of states of infinite-dimensional quantum system $A_1...A_n$ and may take the value $+\infty$ on this set.

A desirable and physically motivated property of any entanglement measure $E$ in the infinite-dimensional case is a lower semicontinuity, which means
that
$$
\liminf_{k\to+\infty} E(\rho_k)\geq E(\rho_0)
$$
for any sequence $\{\rho_k\}$ converging to a state $\rho_0$. This property corresponds to the
natural assumption that the entanglement can not jump up in passing to the limit.

The lower semicontinuity of the entanglement of formation and of the squashed entanglement is proved on the set of states of a bipartite quantum system
having at least one finite marginal entropy \cite{EM,SE}. In \cite{LSE} it is shown that
the $n$-partite squashed entanglement is lower semicontinuous  on the set of states with finite marginal entropies.

Despite the fact that the quantum relative entropy is lower semicontinuous, we have not managed to prove global lower semicontinuity of the relative entropy of entanglement because of the non-compactness of the set of separable states. Nevertheless, by using Theorem \ref{main} in Section 3 and the notion of universal extension of entanglement
monotones considered below we will obtain in Section 4.3 a non-restrictive sufficient condition for local lower semicontinuity of the relative entropy of entanglement and its regularization.
\smallskip

\subsection{Definition of $E_R$ via the universal extension}

Assume that $E$ is a continuous \emph{entanglement monotone} on the set of states of a\break $n$-partite quantum system $A_1...A_n$ composed of finite-dimensional subsystems, i.e.
a continuous function on the set $\S(\H_{A_1...A_n})$ with the following properties (cf.\cite{Vidal,P&V}):
\begin{enumerate}[\textrm{EM}1)]
\item $\{\shs E(\rho)=0\;\}\;\Leftrightarrow\;\{\,\rho\;\,
\textrm{is a separable state}\,\}$;
\item monotonicity under selective unilocal operations:
\begin{equation*}
E(\rho)\geq\sum_{i}p_{i}E(\rho_{i}),\quad
p_{i}=\Tr\shs\Phi_i(\rho),\;\;
\rho_{i}=p_{i}^{-1}\Phi_i(\rho)
\end{equation*}
for any  state $\rho\in\S(\H_{A_1...A_n})$ and any collection $\{\Phi_i\}$ of
unilocal completely positive linear maps such that $\sum_i\Phi_i$ is a
channel.
\item convexity:
\begin{equation*}
E(p\rho_1+(1-p)\rho_2)\leq p E(\rho_1)+(1-p)E(\rho_2)
\end{equation*}
for arbitrary states $\rho_1,\rho_2\in\S(\H_{A_1...A_n})$ and any $p\in(0,1)$.
\end{enumerate}

If $A_1...A_n$ is a $n$-partite quantum system composed of infinite-dimensional subsystems then the
function $E$ is well defined on the set of states $\rho$ in $\S(\H_{A_1...A_n})$ with finite rank marginal states $\,\rho_{A_s}$, $s=\overline{1,n}$.
Thus, one can define the function
\begin{equation}\label{mse-ue}
  \widehat{E}(\rho)\doteq\sup_{P_{1},...,P_{n}}E(P_{1}\otimes...\otimes P_{n}\cdot\rho\cdot P_{1}\otimes...\otimes P_{n})
\end{equation}
on the set $\S(\H_{A_1...A_n})$, where the supremum is over all finite-rank projectors\break $P_{1}\in\B(\H_{A_1})$,...,$P_{n}\in\B(\H_{A_n})$ and it is assumed that
$E(\varrho)=c E(\varrho/c)$ for any nonzero positive operator $\varrho$ in $\T(\H_{A_1..A_n})$ such that $\,\rank\varrho_{A_s}<+\infty$, $s=\overline{1,n}$, where $c=\Tr\varrho$.

The above function $\widehat{E}$ is used in \cite{SE} in the bipartite case, where it is shown that $\widehat{E}$ is a lower semicontinuous  entanglement monotone on $\S(\H_{A_1A_2})$ inheriting many important properties of $E$ (the (sub)additivity for product states, the
monogamy relation, etc). The results
of applying this construction  called the \emph{universal extension} of $E$  to the bipartite squashed entanglement and the entanglement of formation are presented in \cite{SE}.\smallskip

By obvious generalization of the proof of Proposition 5 in \cite{SE} to the $n$-partite case one can prove the following \smallskip

\begin{property}\label{E-prop} \emph{Let $E$ be a continuous entanglement monotone on the set of states of $\,n$-partite quantum system composed of finite-dimensional subsystems. Let
$A_1...A_n$ be an infinite-dimensional $n$-partite quantum system and $\,\S_{\rm f}(\H_{A_1..A_n})$  the set of states $\rho$ in $\,\S(\H_{A_1..A_n})$ with finite rank marginal states $\,\rho_{A_s}$, $s=\overline{1,n}$.}\smallskip

A) \emph{$\widehat{E}$ is a unique lower semicontinuous entanglement monotone on the set $\,\S(\H_{A_1..A_n})$ such that $\widehat{E}(\rho)=E(\rho)$ for any state $\rho$ in $\,\S_{\rm f}(\H_{A_1..A_n})$.}\smallskip

B) \emph{If $\{P^k_{A_s}\}_k\subset\B(\H_{A_s})$, $s=\overline{1,n}$, are arbitrary sequences of finite rank projectors strongly
converging to the unit operators $I_{A_s}$ then
\begin{equation}\label{ue-exp}
\widehat{E}(\rho)=\lim\limits_{k\rightarrow+\infty}E(P^k_{A_1}\otimes...\otimes P^k_{A_n}\cdot\rho\cdot P^k_{A_1}\otimes...\otimes P^k_{A_n})
\end{equation}
for any state $\rho$ in $\,\S(\H_{A_1..A_n})$.}\smallskip

C) \emph{$\widehat{E}$ is the convex closure of $E$ -- the maximal lower semicontinuous convex function on $\,\S(\H_{A_1..A_n})$ not exceeding
the function $E$ on the set $\,\S_{\rm f}(\H_{A_1..A_n})$.}
\end{property}\smallskip

The main advantage of the entanglement monotone $\widehat{E}$ is its global lower semicontinuity. It is this property that is used in \cite{SE} to
show  that $\widehat{E}_{sq}(\rho)=0$ for any separable state $\rho$ in $\S(\H_{A_1A_2})$, while the same
property for the function $E_{sq}$ obtained by direct generalization of the finite-dimensional definition
is hard to prove (until it's proven that $\widehat{E}_{sq}=E_{sq}$) because of the existence of countably-non-decomposable separable states in infinite-dimensional bipartite systems (Remark 10 in \cite{SE}).

Since the relative entropy of entanglement $E_R$ is a continuous entanglement monotone on the set of states of a $n$-partite quantum system composed of finite-dimensional subsystems,
Proposition \ref{E-prop} shows that
\begin{itemize}
  \item the universal extension $\widehat{E}_R$ is a lower semicontinuous  entanglement monotone on the set of states of an infinite-dimensional $n$-partite quantum system;
  \item $\widehat{E}_R(\rho)=E_R(\rho)$ for any state $\rho$ with finite rank marginal states;
  \item $\widehat{E}_R$ is the convex closure of $E_R$ (considered as a function on the set $\,\S_{\rm f}(\H_{A_1..A_n})$), it is explicitly determined for any state $\rho$  by the expression
  (\ref{ue-exp}) with $E=E_R$ via arbitrary sequences $\{P^k_{A_s}\}_k\subset\B(\H_{A_s})$, $s=\overline{1,n}$ of finite rank projectors strongly
converging to the unit operators $I_{A_s}$.
\end{itemize}

The regularization of $\widehat{E}_R$ is defined in the standard way:
\begin{equation}\label{ree-r-def+}
  \widehat{E}^{\infty}_R(\rho)=\lim_{m\rightarrow+\infty}m^{-1}\widehat{E}_R(\rho^{\otimes m}),
\end{equation}
where $\rho^{\otimes m}$ is treated as a state of the $n$-partite quantum system $A^m_1...A^m_n$.\smallskip

For an arbitrary state $\rho$ in $\,\S(\H_{A_1..A_n})$ the  monotonicity  of $E_R$ under selective unilocal operations (which follows from Lemma \ref{LL} at the end of Section 4.3) implies that
$$
\widehat{E}_R(\rho)\leq E_R(\rho)\quad  \textrm{and}\quad \widehat{E}^{\infty}_R(\rho)\leq E^{\infty}_R(\rho)
$$
($E_R$ and $E^{\infty}_R$ are the relative entropy of entanglement obtained by direct generalization of the finite-dimensional definition and its regularization). In the next subsection, we will obtain a simple sufficient condition for equality in these inequalities.

\subsection{Analytical properties of $E_R$ and $E^{\infty}_R$}

The following theorem contains some results about analytical properties of the relative entropy of entanglement
$E_R$ directly defined by formula (\ref{ree-def}) for any state of a $n$-partite infinite-dimensional quantum system and its regularization $E^{\infty}_R$ defined by formula (\ref{ree-r-def}).
It also clarifies relations between the functions $E_R$ and $E^{\infty}_R$, the universal extension $\widehat{E}_R$ defined by formula (\ref{mse-ue}) with $E=E_R$ and its regularization $\widehat{E}^{\infty}_R$  defined in (\ref{ree-r-def+}).\smallskip

\begin{theorem}\label{RE-LS}  \emph{Let
\begin{equation}\label{S-star}
\S_*(\H_{A_1...A_n})=\left\{\shs\rho\in\S(\H_{A_1...A_n})\,|\,\rho_{A_{s}}\!\!\in\S_{\rm\! \textsf{FA}\!}(\H_{A_{s}}) \textrm{ for } \,n-1\, \textrm{ indexes } \shs s\shs\right\}
\end{equation}
be the subset of $\,\S(\H_{A_1...A_n})$ consisting of  states $\rho$
such that the FA-property holds for at least $\shs n-1$
marginal states of $\rho$.}\smallskip

A) \emph{The functions $E_R$ and $E^{\infty}_R$ are finite and lower semicontinuous on the set $\S_*(\H_{A_1...A_n})$.  Moreover,
\begin{equation}\label{LS-r}
\liminf_{k\to+\infty} E^*_R(\rho_k)\geq E^*_R(\rho_0),\quad E^*_R=E_R,E^{\infty}_R,
\end{equation}
for arbitrary sequence $\{\rho_k\}\subset\S(\H_{A_1...A_n})$ converging to a state $\rho_0\in\S_*(\H_{A_1...A_n})$.}\smallskip

B) \emph{The functions $E_R$ and $E^{\infty}_R$ coincide on the set $\,\S_*(\H_{A_1...A_n})$ with the functions $\widehat{E}_R$ and $\widehat{E}^{\infty}_R$ correspondingly.
$\widehat{E}_R$ is the convex closure of $E_R$ (considered as a function on $\S(\H_{A_1...A_n})$).}\smallskip

C) \emph{For any state $\rho\in\S_*(\H_{A_1...A_n})$ the infimum in definition (\ref{ree-def}) of $E_R(\rho)$ can be taken only over all finitely-decomposable
separable states in $\S(\H_{A_1...A_n})$, i.e. states having the form}
\begin{equation}\label{f-d-s}
\sigma=\sum_{i=1}^m p_i\shs \alpha^i_1\otimes...\otimes\alpha^i_n,\quad \alpha^i_s\in\S(\H_{A_s}),\;\; p_i>0,\;\;\sum_{i=1}^m p_i=1,\;\;m<+\infty.
\end{equation}\pagebreak

D) \emph{If the Hamiltonians $H_{\!A_1}$,.., $H_{\!A_{n-1}}$ of subsystems $A_1$,...,$A_{n-1}$ satisfy condition (\ref{H-cond+}) then}\smallskip
\begin{itemize}
  \item \emph{the functions $E_R$ and $E^{\infty}_R$ are  uniformly continuous on the set of states $\rho$ in $\,\S(\H_{A_1...A_n})$ such that  $\sum_{s=1}^{n-1}\Tr H_{A_s}\rho_{A_s}\leq E$ for any $E>E_0^{A_1}+...+E_0^{A_{n-1}}$;}\footnote{$E_0^{A_s}$ is the minimal eigenvalue of $H_{\!A_s}$.}
  \item \emph{the function $E_R$ is uniformly continuous on the set
   $$
   \left\{\Lambda(\rho)\,\left|\,\rho\in\S(\H_{A_1...A_n}),\, \sum_{s=1}^{n-1}\Tr H_{A_s}\rho_{A_s}\leq E\right.\right\}\quad \forall E>E_0^{A_1}+...+E_0^{A_{n-1}},
   $$
   where $\Lambda$  is any positive trace preserving linear  transformation of $\,\T(\H_{A_1...A_n})$ such that $\Lambda(\S_{\rm s}(\H_{A_1...A_n}))\subseteq \S_{\rm s}(\H_{A_1...A_n})$;}
  \item  \emph{the functions $E_R$ and $E^{\infty}_R$ are asymptotically continuous in the following sense (cf.\cite{ESP}):
if $\{\rho_k\}$ and $\{\sigma_k\}$ are any sequences such that
$$
\rho_k,\sigma_k \in\S(\H_{A^k_1...A^k_n}),\,\Tr H_{\!B^{k}}[\rho_k]_{B^{k}},\! \Tr H_{\!B^{k}}[\sigma_k]_{B^{k}}\!\leq kE,\,\forall k,\,\textit{and}\; \lim_{k\to\infty}\|\rho_k-\sigma_k\|_1=0,
$$
where $X^{k}$ denotes $k$ copies of a system $X$, $B=A_1...A_{n-1}$ and $H_{\!B^{k}}$ is the Hamiltonian of $B^{k}$, then}
$$
\lim_{k\to+\infty}\frac{|E^{*}_R(\rho_k)-E^{*}_R(\sigma_k)|}{k}=0, \quad E^*_R=E_R,E^{\infty}_R.
$$
\end{itemize}

E) \emph{The function $E_R$ is $\sigma$-convex on $\S(\H_{A_1...A_n})$, i.e.
$$
E_R(\rho)\leq\sum_{i=1}^{+\infty} p_iE_R(\rho_i),\quad \rho=\sum_{i=1}^{+\infty} p_i\rho_i,
$$
for any countable collection $\{\rho_i\}\subset\S(\H_{A_1...A_n})$ and any probability distribution $\{p_i\}$.\footnote{A convex nonnegative function on the set of quantum states may be not $\sigma$-convex \cite[Ex.1]{EM}.}
The function $E_R$ is $\mu$-integrable \footnote{It means that $E_R$ is measurable w.r.t.
the smallest $\sigma$-algebra on $\S(\H_{A_1...A_n})$ that contains all Borel sets and all the subsets of the zero-$\mu$-measure Borel sets.
The integral in the r.h.s. of (\ref{mu-convex}) is over the corresponding completion of the measure $\mu$ \cite{Rudin}. We can not prove that $E_R$ is a Borel function on $\S(\H_{A_1...A_n})$, since it is defined via the infimum over the non-countable set of separable states.} w.r.t. any Borel probability measure $\mu$ on $\,\S(\H_{A_1...A_n})$ with the barycenter $\,\bar{\rho}(\mu)\doteq\int \!\rho\mu(d\rho)\,$ in $\S_*(\H_{A_1...A_n})$ and}
\begin{equation}\label{mu-convex}
E_R(\bar{\rho}(\mu))\leq\int E_R(\rho)\mu(d\rho).
\end{equation}
\end{theorem}\smallskip

\begin{remark}\label{Th2-C} Theorem \ref{RE-LS}C implies, in particular, that we may define $E_R(\rho)$
for any state $\rho\in\S_*(\H_{A_1...A_n})$ \emph{by ignoring the existence of countably-non-decomposable
separable states} in $\S(\H_{A_1...A_n})$. This assertion is not trivial, since in general there are no reasons to
assume that the infimum in definition (\ref{ree-def}) of $E_R(\rho)$ can be taken over the dense subset of $\S_{\rm s}(\H_{A_1...A_n})$ consisting of countably decomposable
separable states.
\end{remark}\smallskip

\emph{Proof.} A) Let $\rho_0$ be a state in $\S_*(\H_{A_1...A_n})$. Assume that the FA-property holds for the states $[\rho_0]_{A_1}$,...,$[\rho_0]_{A_{n-1}}$.
The finiteness of $E_R(\rho_0)$ and $E^{\infty}_{R}(\rho_0)$ follows from inequality (\ref{ER-UB}), since the FA-property of a state implies
finiteness of its  entropy \cite{FAP}.

For each natural $s$ in $[1,n]$ let $P_r^s$ be the spectral projector of $[\rho_0]_{A_s}$ corresponding to its $r$ maximal eigenvalues (taking the multiplicity into account).
It is mentioned in Section 4.1 that the function $E_R$ belongs to the class $L_n^{n-1}(1,1)$. Hence Theorem \ref{main} implies that
\begin{equation}\label{u-l-r}
\lim_{r\to+\infty} E_R(\rho_r)=E_R(\rho_0),
\end{equation}
where $\rho_r=c_r^{-1}Q_r\rho_0\shs Q_r$,  $\,Q_r=P_r^1\otimes...\otimes P_r^{n}$, $c_r=\Tr Q_r\rho_0$. Since $c_rE_R(\rho_r)$ does not exceed $\widehat{E}_{R}(\rho_0)$  and $c_r$ tends to $1$ as $r\to+\infty$, the above limit relation implies that
\begin{equation}\label{2-eq}
\widehat{E}_{R}(\rho_0)= E_{R}(\rho_0).
\end{equation}

To prove relation (\ref{LS-r}) for $E_R^*=E_{R}$ it suffices to note that $\widehat{E}_{R}$ is a lower semicontinuous function  on $\S(\H_{A_1..A_n})$
not exceeding the function $E_R$ and to use (\ref{2-eq}).

To prove relation (\ref{LS-r}) for $E_R^*=E^{\infty}_{R}$ we apply the arguments from the proof of Theorem \ref{main} to construct for each natural $s$ in $[1,n-1]$ a positive operator
$G_{\!A_s}$ on $\H_{A_s}$ satisfying condition (\ref{H-cond+}) such that
$\Tr G_{\!A_s}[\rho_0]_{A_s}<+\infty$. For each $s$ consider the quantum channel
$$
\Phi^s_r(\varrho)=P_r^s\varrho P_r^s+[\Tr(I_{A_s}-P_r^s)\varrho]\tau_s
$$
from $\T(\H_{A_s})$ to itself, where $P_r^s$ is the spectral projector of $G_{A_s}$ corresponding to its $r$ minimal eigenvalues and $\tau_s$ is a state in $\S(\H_{A_s})$ such that $\Tr G_{A_s}\tau_s=0$ and $P^s_r\tau_sP_r^s=\tau_s$ for all $r$. Since all the output states of $\Phi^s_r$ are supported by the finite-dimensional subspace $P^s_r(\H_{A_s})$, the function
$$
f_r(\rho)=E^{\infty}_{R}(\Phi^1_r\otimes...\otimes\Phi^{n-1}_r\otimes\id_{A_n}(\rho))
$$
is continuous on the set $\S(\H_{A_1...A_n})$ for each $r$ (this follows from the continuity bound (\ref{ER-CB}) for $E^{\infty}_{R}$ obtained independently in Section 4.4 by generalizing the
Winter arguments from \cite{W-CB}). Hence, $f_*=\sup_r f_r$ is a lower semicontinuous function on $\S(\H_{A_1...A_n})$.

Since $\Phi^1_r\otimes...\otimes\Phi^{n-1}_r\otimes\id_{A_n}(\rho_0)$ tends to $\rho_0$ as $r\to+\infty$ and  $\Tr G_{\!A_s}\Phi^s_r([\rho_0]_{A_s})\leq \Tr G_{\!A_s}[\rho_0]_{A_s}<+\infty$ for all $r$ and $s=\overline{1,n-1}$, part D of the theorem implies that
$$
\lim_{r\to+\infty}f_r(\rho_0)=E^{\infty}_R(\rho_0)\quad\textrm{and hence}\quad f_*(\rho_0)=E_R^{\infty}(\rho_0),
$$
as $f_*$ does not exceed the function $E^{\infty}_{R}$ by the monotonicity of $E^{\infty}_{R}$ under local channels.
This implies relation (\ref{LS-r}) for $E_R^*=E^{\infty}_{R}$, since   $f_*$ is a lower semicontinuous function.  \smallskip

B) The properties of the function $\widehat{E}_R$  stated in Section 4.2 show that $\widehat{E}_R$ is the greatest lower semicontinuous convex function on the set $\S(\H_{A_1..A_n})$ not exceeding the function $E_R$.

The arguments from the proof of part A of the theorem show that the functions $E_R$ and $\widehat{E}_R$  coincide on the set $\S_*(\H_{A_1..A_n})$.
The same arguments show that $\widehat{E}_R(\rho^{\otimes m})=E_R(\rho^{\otimes m})$ for any state $\rho$ in $\S_*(\H_{A_1..A_n})$ and arbitrary natural $m$,
since for any such state the state $\rho^{\otimes m}$ belongs to the set $\S_*(\H_{A^m_1..A^m_n})$ by Proposition 1B
in \cite{FAP}. This implies  coincidence of the functions $E^{\infty}_R$ and $\widehat{E}^{\infty}_R$ on the set $\S_*(\H_{A_1..A_n})$.

\smallskip

C) Consider the function
\begin{equation*}
  \widetilde{E}_R(\rho)=\inf_{\sigma\in\S^{\mathrm{f}}_{\mathrm{s}}(\H_{A_1...A_n})}H(\rho\shs\|\shs\sigma),
\end{equation*}
where $\S^{\mathrm{f}}_{\mathrm{s}}(\H_{A_1...A_n})$ is the set of finitely-decomposable separable states in $\S(\H_{A_1...A_n})$. Since
the set $\S^{\mathrm{f}}_{\mathrm{s}}(\H_{A_1...A_n})$ is convex, the joint convexity of the relative entropy and Lemma 6 in \cite{AFM} imply that the function $\widetilde{E}_R$ satisfies inequality (\ref{F-p-1}) with $a_f=1$ and $b_f=0$. Since
$$
 \widetilde{E}_R(\rho)\leq H(\rho\shs\|\shs\rho_{A_1}\otimes...\otimes\rho_{A_n})=I(A_1:...:A_n)_{\rho}\leq 2\sum_{s=1}^{n-1}H(\rho_{A_s}),
$$
where the second inequality follows from (\ref{nMI-UB}), the function $\widetilde{E}_R$ satisfies inequality (\ref{F-p-2}) with $m=n-1$, $c^{-}_f=0$ and $c^{+}_f=2$.
It follows that the function $\widetilde{E}_R$ belongs to the class $L_n^{n-1}(2,1)$.
Hence for any state $\rho_0$ in $\S_*(\H_{A_1...A_n})$ Theorem \ref{main} implies that
$$
\lim_{r\to+\infty}\widetilde{E}_R(\rho_r)=\widetilde{E}_R(\rho_0),
$$
where the notation from the proof of part A of this theorem is used. Since $\widetilde{E}_R(\rho_r)=E_R(\rho_r)$ for any $r$, this limit relation and (\ref{u-l-r})
show that $\widetilde{E}_R(\rho_0)=E_R(\rho_0)$.\smallskip

D) The assertions about uniform continuity of the functions $E_{R}$ and $E^{\infty}_{R}$ follow directly from Propositions \ref{nREE-CB} and \ref{cbp-2} in Section 4.4
below (proved independently).

To prove of the asymptotic continuity of the functions $E_{R}$ and $E^{\infty}_{R}$ note that $F_{H_{\!B^k}}(E)=kF_{H_{\!B}}(E/k)$ and $E_0^{B^k}=kE_0^{B}$  for each $k$ and hence
$\bar{F}_{H_{\!B^k}}(E)=k\bar{F}_{H_{\!B}}(E/k)$, where the notation from Section 4.4 is used. So, Proposition \ref{nREE-CB}  in Section 4.4 with $m=n-1$ implies that
\begin{equation}\label{SE-ucb-k}
    \frac{|E_R^*(\rho_k)-E_R^*(\sigma_k)|}{k}\leq \displaystyle \sqrt{2\varepsilon_k}\bar{F}_{H_{\!B}}\!\!\left(\bar{E}/\varepsilon_k\right)+(1/k)g(\sqrt{2\varepsilon_k}),\;\; E_R^*=E^{\infty}_{R},E_R,
\end{equation}
where $\varepsilon_k=\frac{1}{2}\|\shs\rho_k-\sigma_k\|_1$ and $\bar{E}=E-E_0^B$. Since the sequence $\{\varepsilon_k\}$ is vanishing by the condition
and $\bar{F}_{H_{\!B}}(E)$ is $o(\sqrt{E})$ as $E\to+\infty$ by Lemma 1 in \cite{CBM},
the r.h.s. of (\ref{SE-ucb-k}) tends to zero as $k\to+\infty$. \smallskip

E) For arbitrary $\varepsilon>0$ and each $i$ let $\sigma_i$ be a separable state
such that $E_R(\rho_i)\geq H(\rho_i\|\shs\sigma_i)-\varepsilon$. Let $\rho_k=c_k^{-1}\sum_{i=1}^k p_i\rho_i$ and $\sigma_k=c_k^{-1}\sum_{i=1}^k p_i\sigma_i$, where $c_k=\sum_{i=1}^k p_i$.
By the joint convexity of the relative entropy we have
$$
H(\rho_k\|\shs\sigma_k)\leq c_k^{-1}\sum_{i=1}^k p_i H(\rho_i\|\shs\sigma_i).
$$
By using the lower semicontinuity of the relative entropy we obtain
$$
E_R(\rho)\leq H(\rho\shs\|\shs\sigma)\leq \sum_{i=1}^{+\infty} p_i H(\rho_i\|\shs\sigma_i)\leq \sum_{i=1}^{+\infty}p_iE_R(\rho_i)+\varepsilon,
$$
where $\sigma=\sum_{i=1}^{+\infty}p_i\sigma_i$ and the first inequality follows from the separability of $\sigma$. \smallskip

Let $\mu$ be a Borel probability measure on $\S(\H_{A_1...A_n})$ with the barycenter $\rho=\bar{\rho}(\mu)$ in $\S_*(\H_{A_1...A_n})$.
Assume that the FA-property holds for the states $\rho_{A_1}$,...,$\rho_{A_{n-1}}$. Following the proof of Theorem \ref{main}
for each natural $s$ in $[1,n-1]$ we construct a positive operator
$G_{\!A_s}$ on $\H_{A_s}$ satisfying condition (\ref{H-cond+}) such that
$\Tr G_{\!A_s}\rho_{A_s}<+\infty$.

By part D of the theorem the function $E_R$ is continuous on the closed subset $\C_{E}$  of $\S(\H_{A_1,..,A_n})$ consisting of states  $\rho$ such that $\sum_{s=1}^{n-1}\Tr G_{\!A_s}\rho_{A_s}\leq E$ for any $E>0$. It follows that the function $f_E$ coinciding with $E_R$ on the set $\C_{E}$ and equal
to $\,+\infty\,$ on the set $\S(\H_{A_1,..,A_n})\setminus\C_{E}$ is a Borel function on $\S(\H_{A_1,..,A_n})$ for each $E>0$.

Since
$$
\int\left[\sum_{s=1}^{n-1}\Tr G_{\!A_s}\varrho_{A_s}\right] \mu(d\varrho)=\sum_{s=1}^{n-1}\Tr G_{\!A_s}\rho_{A_s}<+\infty,
$$
we have $\mu(\S(\H_{A_1...A_n})\setminus\C_{*})=0$, where $\C_{*}=\cup_{E>0}\C_E$.
Hence the function $E_R$
is\break $\mu$-integrable as it coincides with the Borel function $f_*=\inf_{E>0} f_E$
on the set $\C_{*}$.

The function $\widehat{E}_R$ defined in (\ref{mse-ue}) with $E=E_R$ is a convex and lower semicontinuous
function on $\S(\H_{A_1...A_n})$ not exceeding the function $E_R$. Hence, by using the validity of Jensen's inequality for $\widehat{E}_R$ (cf. \cite[the Appendix]{EM}) we obtain
$$
E_R(\bar{\rho}(\mu))=\widehat{E}_R(\bar{\rho}(\mu))\leq\int \widehat{E}_R(\rho)\mu(d\rho)\leq\int E_R(\rho)\mu(d\rho),
$$
where the equality follows from part B of the theorem. $\square$ \medskip

In the following corollary we present several  sufficient conditions for convergence of the relative entropy of entanglement and its regularization used below.\smallskip
\begin{corollary}\label{RE-LS-c+}  \emph{Let $\{\rho_k\}$ be a sequence of states in $\S(\H_{A_1...A_n})$ converging to a state $\rho_0$
from the set $\,\S_*(\H_{A_1...A_n})$ defined in (\ref{S-star}). Then the relation
\begin{equation}\label{CS-r+}
\lim_{k\to+\infty} E_R(\rho_k)=E_R(\rho_0)<+\infty
\end{equation}\vspace{-5pt}
holds provided that one of the following conditions is valid:}
\begin{enumerate}[a)]
 \item \emph{all the states $\rho_k$
 are obtained from  the state $\rho_0$ by LOCC;}\vspace{-5pt}
 \item \emph{$\rho_k=\Phi_k(\rho_0)/\Tr\Phi_k(\rho_0)$, where $\Phi_k$ is a positive trace-non-increasing linear
 transformation of $\,\T(\H_{A_1...A_n})$ such that \footnote{$\mathrm{conv}\{0,\S_{\rm s}(\H_{A_1...A_n})\}$ is the set of all operators of the form $\,c\rho$, where $\rho\in\S_{\rm s}(\H_{A_1...A_n})$ and $c\leq 1$. Condition (\ref{Phi-cond}) does not imply that $\Phi_k$ is a separable transformation (Definition 3.38 in \cite{Wilde-new})), see Example \ref{CS-r++r} below.}
 \begin{equation}\label{Phi-cond}
 \Phi_k(\S_{\rm s}(\H_{A_1...A_n}))\subseteq \mathrm{conv}\{0,\S_{\rm s}(\H_{A_1...A_n})\}
 \end{equation}
 for each $k$ and $\Tr\Phi_k(\rho_0)$ tends to $1$ as $k\to +\infty$;}\vspace{-5pt}
 \item \emph{$c_k\rho_k\leq \sigma_k$ for all $\shs k$, where $\{c_k\}$ is a sequence of positive numbers tending to $1$ and $\{\sigma_k\}$  is a  sequence of states in $\S(\H_{A_1...A_n})$ converging to the state $\rho_0$ such that}
$$
  \lim_{k\to+\infty} E_R(\sigma_k)=E_R(\rho_0).
$$
\end{enumerate}
\emph{The above condition a) implies  that
\begin{equation}\label{CS-r++}
\lim_{k\to+\infty} E^{\infty}_R(\rho_k)=E^{\infty}_R(\rho_0)<+\infty.
\end{equation}
The above condition b) also implies relation (\ref{CS-r++}) provided that $\,\Phi_k$ is a completely positive trace-non-increasing linear
transformation of $\T(\H_{A_1...A_n})$ such that the map $\Phi^{\otimes m}_k$ satisfies the $m$-partite version of condition (\ref{Phi-cond}) for each natural $m$ and all $\,k$.}\footnote{It means that $\Phi^{\otimes m}_k(\S_{\rm s}(\H_{A^m_1...A^m_n}))\subseteq \mathrm{conv}\{0,\S_{\rm s}(\H_{A^m_1...A^m_n})\}$, where $A^m_s$ denotes the composition of $\,m$ copies of the system $A^m_s$.}
\end{corollary}\smallskip

\emph{Proof.} a) The LOCC monotonicity of $E_R$ and $E^{\infty}_R$ implies that $E^*_R(\rho_k)\leq E^*_R(\rho_0)$, $E^{*}_R=E_R,E^{\infty}_R$, for all $k$. So, relations (\ref{CS-r+}) and (\ref{CS-r++}) follow from Theorem \ref{RE-LS}A.

b) Lemma \ref{LL} below  implies that $\Tr\Phi_k(\rho_0)E_R(\rho_k)\leq E_R(\rho_0)$ for all $k$. So, in this case (\ref{CS-r+})
also follows from Theorem \ref{RE-LS}A.

If all the maps $\,\Phi_k$  are completely positive and the maps $\Phi^{\otimes m}_k$ satisfy the\break $m$-partite  version  of condition (\ref{Phi-cond}) for each natural $m$ then
Lemma \ref{LL} below  implies that $\Tr\Phi^{\otimes m}_k(\rho^{\otimes m}_0) E_R(\rho^{\otimes m}_k)\leq E_R(\rho^{\otimes m}_0)$ for all $k$. Proposition 1B in \cite{FAP} implies that the state $\rho_0^{\otimes m}$ belongs to the set $\S_*(\H_{A^m_1...A^m_n})$. So, it follows from the above inequality and Theorem \ref{RE-LS}A that
\begin{equation*}
\lim_{k\to+\infty} E_R(\rho^{\otimes m}_k)=E_R(\rho^{\otimes m}_0)\quad \forall m.
\end{equation*}
Since $E_R^{\infty}(\rho_k)=\inf_m m^{-1}E_R(\rho_k^{\otimes m})$, $k=0,1,2,..$, the validity of the last relation for all $m$ shows that
\begin{equation*}
\limsup_{k\to+\infty} E^{\infty}_R(\rho_k)\leq E^{\infty}_R(\rho_0).
\end{equation*}
This limit relation and Theorem \ref{RE-LS}A imply (\ref{CS-r++}).

\smallskip

c) Since the function $E_R$ satisfies inequality (\ref{F-p-1}) with $a_f=1$ and $b_f=0$, we have
$$
E_R(\sigma_k)\geq c_k E_R(\rho_k)+(1-c_k)E_R((\sigma_k-c_k\rho_k)/(1-c_k))-h_2(c_k)\geq c_k E_R(\rho_k)-h_2(c_k)\quad \forall k.
$$
It follows that
$$
\limsup_{k\to+\infty} E_R(\rho_k)\leq \lim_{k\to+\infty} E_R(\sigma_k)=E_R(\rho_0).
$$
This relation and Theorem \ref{RE-LS}A imply (\ref{CS-r+}). $\square$
\medskip

Consider simple applications of continuity conditions in Corollary \ref{RE-LS-c+}. \smallskip

\begin{example}\label{CS-r++r} Let $\rho_0$ be a state from the set $\,\S_*(\H_{A_1...A_n})$ and $\{\sigma_k\}$ an arbitrary sequence of separable states
in $\,\S(\H_{A_1...A_n})$. Then
\begin{equation}\label{CS-r+ex}
\lim_{k\to+\infty} E^*_R((1-p_k)\rho_0+p_k\sigma_k)=E^*_R(\rho_0),\quad E^*_R=E_R,E_R^{\infty},
\end{equation}
for any vanishing sequence $\{p_k\}\in[0,1]$. Indeed, let $\Phi_k(\rho)=(1-p_k)\rho+p_k\sigma_k$ be a channel from
$\,\T(\H_{A_1...A_n})$ to itself. It is easy to see that the channel $\Phi_k^{\otimes m}$ satisfies the $m$-partite version of condition (\ref{Phi-cond}) for each natural $m$ and all $\,k$.
So, both limit relations in (\ref{CS-r+ex}) follow from  condition b) in Corollary \ref{RE-LS-c+}.
\end{example}\smallskip

\begin{example}\label{CS-r++r+} Let $\rho_0$ be a state from the set $\,\S_*(\H_{A_1...A_n})$ and $\{\rho_k\}$ a sequence of states
converging to the state $\rho_0$ such that $c_k\rho_k\leq \rho_0$ for all $k$, where $\{c_k\}$ is a sequence of positive numbers tending to $1$. Then
\begin{equation}\label{CS-r+ex+}
\lim_{k\to+\infty} E^*_R(\rho_k)=E^*_R(\rho_0),\quad E^*_R=E_R,E_R^{\infty}.
\end{equation}
Indeed, relation (\ref{CS-r+ex+}) with $E_R^*=E_R$ directly follows from condition c) in Corollary \ref{RE-LS-c+}.
Since  $\,c^m_k\rho^{\otimes m}_k\leq \rho_0^{\otimes m}\,$ for each natural $m$ and the state $\rho_0^{\otimes m}$ belongs to the set $\S_*(\H_{A^m_1...A^m_n})$ by Proposition 1B in \cite{FAP}, this condition also implies that
\begin{equation*}
\lim_{k\to+\infty} E_R(\rho^{\otimes m}_k)=E_R(\rho^{\otimes m}_0)\quad \forall m.
\end{equation*}
Since $E_R^{\infty}(\rho_k)=\inf_m m^{-1}E_R(\rho_k^{\otimes m})$, $k=0,1,2,..$, the validity of the last relation for all $m$ shows that
\begin{equation*}
\limsup_{k\to+\infty} E^{\infty}_R(\rho_k)\leq E^{\infty}_R(\rho_0).
\end{equation*}
This limit relation and Theorem \ref{RE-LS}A imply (\ref{CS-r+ex+}) with $E_R^*=E_R^{\infty}$.
\end{example}\smallskip

Continuity condition c) in Corollary \ref{RE-LS-c+} will be essentially used in the proof of Proposition \ref{REE-FDA} in Section 4.\smallskip

\begin{lemma}\label{LL}
\emph{Let $\{\Phi_i\}$ be a collection of positive linear maps from $\T(\H)$ to itself such that the map
$\sum_i\Phi_i$ is trace preserving. Let $\rho$ and $\sigma$ be any states in $\S(\H)$. Then
$$
\sum_i p_iH(\rho_i\|\shs\sigma_i)\leq H(\rho\shs\|\shs\sigma),
$$
where $p_i=\Tr\Phi_i(\rho)$, $\rho_i=p_i^{-1}\Phi_i(\rho)$, $q_i=\Tr\Phi_i(\sigma)\shs$ and $\,\sigma_i=q_i^{-1}\Phi_i(\sigma)$.}
\end{lemma}\smallskip

\emph{Proof.} Consider the trace preserving positive linear map
$$
\T(\H)\ni\rho\mapsto\widetilde{\Phi}(\rho)=\sum_{i}\Phi_i(\rho)\otimes |i\rangle\langle i|\in\T(\H\otimes\H_R),
$$
where $\{|i\rangle\}$ is a basis in appropriate Hilbert space $\H_R$.  By using  the monotonicity of the
relative entropy under action of the map $\widetilde{\Phi}$ (proved in \cite{Reeb}) and the properties of the Lindblad extension (\ref{qre-L}) of the
relative entropy (presented in \cite{L-2}) we obtain
$$
\begin{array}{c}
 \displaystyle H(\rho\shs\|\shs\sigma)\geq  H(\widetilde{\Phi}(\rho)\shs\|\shs\widetilde{\Phi}(\sigma))=\sum_i  H(p_i\rho_i\shs\|\shs q_i\sigma_i)=\sum_i p_i H(\rho_i\shs\|\shs q_i\sigma_i/p_i)\;\\\\
 \displaystyle=
 \sum_i p_i H(\rho_i\shs\|\shs\sigma_i)+D(\{p_i\}\|\{q_i\})\geq  \sum_i p_i H(\rho_i\shs\|\shs\sigma_i),
\end{array}
$$
where the last inequality follows from the nonnegativity of the Kullback–Leibler divergence $D(\{p_i\}\|\{q_i\})$ between the probability distributions $\{p_i\}$ and $\{q_i\}$. $\square$ \medskip

\subsection{Finite-dimensional approximation of $E_R$ and $E_R^{\infty}$}

Many results describing properties of the relative entropy of entanglement and its regularization in finite-dimensional multipartite quantum systems
remain valid in the infinite-dimensional case under some additional conditions. A simple way to establish validity of these
results is given by the following\smallskip

\begin{property}\label{REE-FDA}  \emph{Let $\rho$ be a state from the set $\,\S_*(\H_{A_1...A_n})$ defined in (\ref{S-star}).\footnote{$\,\S_*(\H_{A_1...A_n})$ is the subset of $\S(\H_{A_1...A_n})$ consisting of  states $\rho$
such that the FA-property holds for at least $\shs n-1$
marginal states of $\rho$.} Let $\rho_k=Q_k\rho\shs Q_k [\Tr Q_k\rho\shs]^{-1}$, $Q_k=P_k^{1}\otimes...\otimes P_k^{n}$,
where $\{P_k^1\}\subset\B(\H_{A_1})$,..., $\{P_k^n\}\subset\B(\H_{A_n})$ are arbitrary sequences of projectors strongly converging to the unit operators
$I_{A_1}$,...,$I_{A_n}$ correspondingly.  Then
\begin{equation}\label{FDA-LR}
E^*_R(\rho_{A_1...A_m})=\lim_{k\to+\infty}E^*_R([\rho_k]_{A_1...A_m}),\quad E_R^*=E_R,E_R^{\infty},
\end{equation}
for any $\,m=2,3,...,n$, where $E_R$ and $E_R^{\infty}$ denote, respectively, the relative entropy of entanglement and its regularization of a state of the system $A_1...A_m$.}\smallskip
\end{property}

\emph{Proof.} For each $k$ let
$$
\sigma_k=c^{-1}_{k,m}Q_{k,m}\shs\rho_{A_1...A_m}\shs Q_{k,m},\quad c_{k,m}=\Tr Q_{k,m}\shs\rho_{A_1...A_m},\quad  Q_{k,m}=P_k^{1}\otimes...\otimes P_k^{m},
$$
be a state of the system $A_1...A_m$. Then for any natural $u$ we have
$$
\sigma_k^{\otimes u}=c^{-u}_{k,m}[Q_{k,m}]^{\otimes u}\rho^{\otimes u}_{A_1...A_m}\shs [Q_{k,m}]^{\otimes u}.
$$
By the monotonicity of $E_R$ under selective unilocal operations we have
$$
c^u_{k,m} E_R(\sigma_k^{\otimes u})\leq E_R(\rho^{\otimes u}_{A_1...A_m}).
$$
Since the state $\rho_{A_1...A_m}$ belongs to the set $\S_*(\H_{A_1...A_m})$, Proposition 1B in \cite{FAP} implies that the state $\rho^{\otimes u}_{A_1...A_m}$ belongs to the set $\S_*(\H_{A^u_1...A^u_m})$. So, the above inequality and Theorem \ref{RE-LS}A show that
$$
\lim_{k\to+\infty}E_R(\sigma^{\otimes u}_k)=E_R(\rho^{\otimes u}_{A_1...A_m})<+\infty,\quad u=1,2,...
$$
Since $c^u_k[\rho_k]^{\otimes u}_{A_1...A_m}\leq\sigma^{\otimes u}_k$ for each natural $u$, where $c_k=\Tr Q_k\rho/\Tr Q_{k,m}\shs\rho_{A_1...A_m}$ is a number tending to $1$ as $k\to+\infty$, the last limit relation implies, by Corollary \ref{RE-LS-c+} with condition c), that
$$
\lim_{k\to+\infty}E_R([\rho_k]^{\otimes u}_{A_1...A_m})=E_R(\rho^{\otimes u}_{A_1...A_m}),\quad u=1,2,...
$$
This relation with $u=1$ means (\ref{FDA-LR}) for $E_R^*=E_R$, while its validity for all $u$ shows that
$$
\limsup_{k\to+\infty}E^{\infty}_R([\rho_k]_{A_1...A_m})\leq E^{\infty}_R(\rho_{A_1...A_m}).
$$
Since the state $\rho_{A_1...A_m}$ belongs to the set $\S_*(\H_{A_1...A_m})$, this relation and Theorem \ref{RE-LS}A
imply (\ref{FDA-LR}) for $E_R^*=E^{\infty}_R$. $\square$
\smallskip

\begin{remark}\label{REE-FDA-r}  If $\rho$ is a state in $\,\S_*(\H_{A_1...A_n})$ and $\{\rho_k\}$ is the sequence defined in Proposition \ref{REE-FDA} by means
of the sequences $\{P_r^1\}$,...,$\{P_r^n\}$ consisting of the spectral projectors of the states $\rho_{A_1}$,...,$\rho_{A_n}$ corresponding to their $r$ maximal eigenvalues
then  by using Proposition \ref{cbp-2} in Section 4.4 one can  show that
\begin{equation*}
E_R(\Lambda([\rho_k]_{A_1...A_m}))\mathop{\rightrightarrows}\limits_{\Lambda}E_R(\Lambda(\rho_{A_1...A_m}))\quad\textrm{as}\quad k\to+\infty,
\end{equation*}
where $\,\mathop{\rightrightarrows}\limits_{\Lambda}\,$ denotes the uniform convergence on the set of all positive trace-non-increasing linear
transformations of $\T(\H_{A_1...A_n})$ satisfying condition (\ref{L-cond}).
\end{remark}\smallskip

Below we consider examples of using Proposition \ref{REE-FDA}.\smallskip

\begin{example}\label{one}
Lemma 5 in \cite{PVP} and the additivity of the entropy imply that
\begin{equation}\label{LB-1}
E_R^{\infty}(\rho)\geq -H(A_i|A_j)_{\rho},\quad (i,j)=(1,2),(2,1),
\end{equation}
for any state $\rho$ of a bipartite finite-dimensional quantum system $A_1A_2$.\smallskip

Proposition \ref{REE-FDA} allows to show that the inequalities in (\ref{LB-1}) remain valid
for any state $\rho$ of a bipartite infinite-dimensional system provided that
the state $\rho_{A_i}$ has the\break FA-property\footnote{This assumption implies, by Theorem 1 in \cite{FAP}, the finiteness
of the extended conditional entropy $H(A_i|A_j)_{\rho}$ defined in (\ref{ce-ext}).} and $H(\cdot|\cdot)$ is the extended conditional entropy defined in (\ref{ce-ext}).
Indeed, it suffices to show that
$$
H(A_1|A_2)_{\rho}=\lim_{k\to+\infty}H(A_1|A_2)_{\rho_k},\quad \rho_k=Q_k\rho\shs Q_k [\Tr Q_k\rho\shs]^{-1},\quad Q_k=P_k^{1}\otimes P_k^{2},
$$
where $P_k^s$ is the spectral projector of $\rho_{A_s}$ corresponding to its $r$ maximal eigenvalues, $s=1,2$.
This can be done by using Theorem \ref{main}, since the function $\rho\mapsto H(A_1|A_2)_{\rho}$
belongs to the class $L_2^1(2,1)$ (see Section 3).

Note that direct proof of (\ref{LB-1}) in the infinite-dimensional case
requires technical efforts (especially, if $H(\rho)=H(\rho_{A_j})=+\infty$).
\end{example}\smallskip

\begin{example}\label{two} It is shown in \cite{ERUB} that
\begin{equation}\label{LB-2}
E_R^*(\rho)\geq E_R^*(\rho_{A_iA_j})+H(\rho_{A_iA_j}),\;\; E_R^*=E_R,E_R^{\infty},\;\;(i,j)=(1,2),(2,3),(3,1),
\end{equation}
for any pure state $\rho$ in a tripartite finite-dimensional quantum system $A_1A_2A_3$.

Proposition \ref{REE-FDA} allows to show that all the inequalities in (\ref{LB-2}) remain valid
for any pure state $\rho$ of a tripartite infinite-dimensional system provided that
any two of the  states $\rho_{A_1}$, $\rho_{A_2}$ and $\rho_{A_3}$ have the FA-property.\footnote{This assumption and the purity of the state $\rho$ imply, by Theorem 1 in \cite{FAP}, the finiteness
of all the entropies $H(\rho_{A_iA_j})$ involved in (\ref{LB-2}).}
Indeed, it suffices to show that
$$
H(\rho_{A_1A_2})=\lim_{k\to+\infty}H([\rho_k]_{A_1A_2}),\quad \rho_k=Q_k\rho\shs Q_k [\Tr Q_k\rho\shs]^{-1},\quad Q_k=P_k^{1}\otimes P_k^{2}\otimes P_k^{3},
$$
where $P_k^s$ is the spectral projector of $\rho_{A_s}$ corresponding to its $r$ maximal eigenvalues, $s=1,2,3$.
This can be done by noting that $H(\rho_{A_1A_2})=H(\rho_{A_3})$ and $H([\rho_k]_{A_1A_2})=H([\rho_k]_{A_3})$ as the states $\rho$ and $\rho_k$ are pure. Since $c_k[\rho_k]_{A_3}\leq \rho_{A_3}$ for each $k$, where $c_k$ is a number
tending to $1$ as $k\to+\infty$, the required limit relation can be proved easily by using concavity and lower semicontinuity of the entropy.
\end{example}

\section{On energy-constrained versions of $E_R$}

Dealing with infinite-dimensional quantum systems we have to take into account the existence of quantum states with infinite energy which can not be
produced in a physical experiment. This motivates an idea to define the relative entropy of entanglement of a state $\rho$
with finite energy by formula (\ref{ree-def}) in which the infimum is taken over all separable states with the energy not exceeding some (sufficiently large) bound $E$ \cite{DNWL}.
This gives the following energy-constrained version of the relative entropy of entanglement
\begin{equation}\label{ree-H-E-def}
  E^{H_{\!A^n}}_R(\rho\shs|E)=\inf_{\sigma\in\S_{\mathrm{s}}(\H_{A_1...A_n}),\Tr  H_{\!A^n}\sigma\leq E}H(\rho\shs\|\shs\sigma),\quad
\end{equation}
where $H_{\!A^n}$ is the Hamiltonian of system $A^n\doteq A_1...A_n$. This definition looks attractive from mathematical point of view, since
the subset of $\S_{\mathrm{s}}(\H_{A_1...A_n})$ satisfying the inequality $\Tr H_{\!A^n}\sigma\leq E$ is compact provided that the Hamiltonian $H_{\!A^n}$
has discrete spectrum of finite multiplicity. This implies, by the lower semicontinuity of the relative entropy, that the infimum in (\ref{ree-H-E-def}) is attainable
(this advantage was exploited in \cite{DNWL}).

An obvious drawback of definition (\ref{ree-H-E-def})  is its dependance of the bound $E$. Note also that
$E^{H_{\!A^n}}_R(\rho\shs|E)\neq0$ for any separable state $\rho$ such that $\Tr H_{\!A^n}\rho>E$ and that
the monotonicity of $E^{H_{\!A^n}}_R(\rho\shs|E)$ under local operations can be shown only under the appropriate restrictions on
the energy amplification factors of such operations.

A less restrictive way to take the energy constraints into account is to take the infimum in
formula (\ref{ree-def})  over all separable states with finite energy, i.e. to consider the following quantity
\begin{equation}\label{ree-H-def}
  E^{H_{\!A^n}}_R(\rho)=\inf_{\sigma\in\S_{\mathrm{s}}(\H_{A_1...A_n}),\Tr H_{\!A^n}\sigma<+\infty}H(\rho\shs\|\shs\sigma)=\lim_{E\rightarrow+\infty}E^{H_{\!A^n}}_R(\rho\shs|E),
\end{equation}
where the limit can be replaced by the infimum over all $E>0$, since the function $E\mapsto E^{H_{\!A^n}}_R(\rho\shs|E)$ is non-increasing. If the Hamiltonian $H_{\!A^n}$ of the system $A_1...A_n$ has the form
\begin{equation}\label{Hn}
H_{\!A^n}=H_{A_1}\otimes I_{A_2}\otimes...\otimes I_{A_n}+\cdots+I_{A_1}\otimes... \otimes I_{A_{n-1}}\otimes H_{A_n}
\end{equation}
then the set of separable states with finite energy is dense in $\S_{\mathrm{s}}(\H_{A_1...A_n})$. So, in this case it is reasonable
to assume the coincidence of $E^{H_{\!A^n}}_R(\rho)$ and $E_R(\rho)$ for any state $\rho$ with finite energy.\footnote{This coincidence is not obvious, since the infima of a lower semicontinuous function
over a closed set and over a dense subset if this set may be different.} In the following proposition  we establish the validity of this assumption under the particular condition
on the Hamiltonians $H_{A_1}$,...,$H_{A_n}$.\smallskip

\begin{property}\label{EC-REE} \emph{Let $H_{\!A^n}$ be the Hamiltonian of a composite system $A_1...A_n$ expressed by formula (\ref{Hn}) via the Hamiltonians $H_{A_1}$,...,$H_{A_n}$ of the subsystems $A_1$,...,$A_n$. If at least $\,n-1\,$ of them satisfy condition (\ref{H-cond+}) then
\begin{equation}\label{E-conject}
  E^{H_{\!A^n}}_R(\rho)=E_R(\rho)
\end{equation}
for any state $\rho$ such that $\Tr H_{\!A^n}\rho=\sum_{s=1}^n\Tr H_{A_s}\rho_{A_s}<+\infty$. Moreover,
for any such  state $\rho$ the infimum in definition (\ref{ree-def}) of $E_R(\rho)$ can be taken only over all finitely-decomposable
separable states $\sigma$ (i.e. the states having form (\ref{f-d-s})) such that  $\Tr H_{\!A^n}\sigma=\sum_{s=1}^n\Tr H_{A_s}\sigma_{A_s}<+\infty$.}\smallskip
\end{property}\smallskip

\emph{Proof.} Assume that the Hamiltonians $H_{A_1}$,...,$H_{A_{n-1}}$ satisfy condition (\ref{H-cond+}). Let $\S_0$ be the convex subset of
$\S(\H_{A_1...A_n})$ consisting of all states $\rho$ with finite value of $\Tr H_{\!A^n}\rho=\sum_{s=1}^n\Tr H_{A_s}\rho$. On the set $\S_0$
consider the function
\begin{equation*}
  \widetilde{E}^{H_{\!A^n}}_R(\rho)=\inf_{\sigma\in\S^{\mathrm{f}}_{\mathrm{s}}(\H_{A_1...A_n})\cap\S_0}H(\rho\shs\|\shs\sigma),
\end{equation*}
where $\S^{\mathrm{f}}_{\mathrm{s}}(\H_{A_1...A_n})$ is the set of finitely-decomposable separable states in $\S(\H_{A_1...A_n})$. Since
the set $\S^{\mathrm{f}}_{\mathrm{s}}(\H_{A_1...A_n})\cap\S_0$ is convex, the joint convexity of the relative entropy and Lemma 6 in \cite{AFM} imply that the function  $\widetilde{E}^{H_{\!A^n}}_R$ satisfies inequality (\ref{F-p-1}) with $a_f=1$ and $b_f=0$ on the set $\S_0$. Since
$$
 \widetilde{E}^{H_{\!A^n}}_R(\rho)\leq H(\rho\shs\|\shs\rho_{A_1}\otimes...\otimes\rho_{A_n})=I(A_1:...:A_n)_{\rho}\leq 2\sum_{s=1}^{n-1}H(\rho_{A_s})\quad \forall \rho\in\S_0,
$$
where the second inequality follows from (\ref{nMI-UB}), the function $\widetilde{E}^{H_{\!A^n}}_R$ satisfies inequality (\ref{F-p-2}) with $m=n-1$, $c^{-}_f=0$ and $c^{+}_f=2$ on the set $\S_0$.
In terms of Remark 3 in \cite{CBM} it means that the function $\widetilde{E}^{H_{\!A^n}}_R$ belongs to the class $L_n^{n-1}(2,1|\shs\S_0)$. It is clear that $\,E_R(\rho)\leq E^{H_{\!A^n}}_R(\rho)\leq\widetilde{E}^{H_{\!A^n}}_R(\rho)\,$ for any state $\rho$ in $\S_0$.

For each natural $s$ in $[1,n-1]$ let $P_r^s$ be the spectral projector of the operator $H_{A_s}$ corresponding to its $r$ minimal eigenvalues (taking the multiplicity into account).
Let $P_r^n=\sum_{i=1}^r|i\rangle\langle i|$, where $\{|i\rangle\}$ is a basis in $\H_{A_n}$ such that $\langle  i|H_{A_n}|i\rangle<+\infty$ for all $i$.

By the proof of Theorem 1 in \cite{AFM} the set $\S_0$ has the invariance property stated in Remark 3 in \cite{CBM}. By this remark Theorem 1 in \cite{CBM} is generalized
to all the functions from the class $L_n^{n-1}(2,1|\shs\S_0)$ containing the function $\widetilde{E}^{H_{\!A^n}}_R$. Since the function $E_R$ belongs to the class  $L_n^{n-1}(1,1)$,  Theorem 1 in \cite{CBM}, its generalization mentioned before and the arguments from the proof of Theorem \ref{main} in Section 3 show that
\begin{equation}\label{u-l-r++}
\lim_{r\to+\infty} E_R(\rho_r)=E_R(\rho)\quad \textrm{and} \quad \lim_{r\to+\infty} \widetilde{E}^{H_{\!A^n}}_R(\rho_r)=\widetilde{E}^{H_{\!A^n}}_R(\rho),\quad \forall\rho\in\S_0,
\end{equation}
where $\rho_r=c_r^{-1}Q_r\rho\shs Q_r$,  $\,Q_r=P_r^1\otimes...\otimes P_r^{n}$, $c_r=\Tr Q_r\rho$. Hence, to prove that $\widetilde{E}^{H_{\!A^n}}_R(\rho)=E^{H_{\!A^n}}_R(\rho)=E_R(\rho)$ it suffices to show that
$\widetilde{E}^{H_{\!A^n}}_R(\rho_r)=E_R(\rho_r)$ for any $r$. This can be down by using Lemma \ref{loc-l} below and by noting that $\Tr H_{\!A^n}\sigma=\sum_{s=1}^n\Tr H_{A_s}\sigma_{A_s}<+\infty$ for any separable state
$\sigma$ supported by the subspace $P_r^1\otimes...\otimes P_r^{n}(\H_{A_1...A_n})$. $\square$
\smallskip

\begin{lemma}\label{loc-l}  \emph{For an arbitrary state $\rho$ in $\S(\H_{A_1...A_n})$ the infimum in (\ref{ree-def})
can be taken over the set of all separable states in $\,\S(\H_{A_1...A_n})$ supported by the subspace
$\H^1_{\rho}\otimes...\otimes \H^n_{\rho}$, where  $\H^s_{\rho}=\supp \rho_{A_s}$, $s=\overline{1,n}$.}
\end{lemma}\smallskip

\emph{Proof.} Consider the channel
$$
\Phi(\varrho)=Q\varrho\shs Q+[\Tr(I_{A_1...A_n}-Q)\varrho]\tau,\quad  Q=P_{1}\otimes...\otimes P_{n},
$$
where $P_s$ is the projector on the subspace $\H^s_{\rho}$ and $\tau$ is any separable state in  $\S(\H_{A_1...A_n})$ supported by the subspace
$\H^1_{\rho}\otimes...\otimes \H^n_{\rho}$. By monotonicity of the relative entropy we have
$$
H(\rho\shs\|\Phi(\sigma))=H(\Phi(\rho)\|\Phi(\sigma))\leq H(\rho\shs\|\shs\sigma)
$$
for any separable state $\sigma$ in $\S(\H_{A_1...A_n})$. Since $\Phi(\sigma)$ is
a separable state supported by the subspace
$\H^1_{\rho}\otimes...\otimes \H^n_{\rho}$, the above inequality implies the assertion of the lemma.  $\square$

\bigskip

\section*{Appendix}

The following lemma is a $n$-partite generalization of the observation in \cite{V&P}.\smallskip

\begin{lemma}\label{Omega}  \emph{For any pure state $\omega$ in $\S(\H_{A_1...A_n})$ there is a countably decomposable
separable state $\sigma$ in $\S(\H_{A_1...A_n})$ such that $\sigma_{A_s}=\omega_{A_s}$ for $s=\overline{1,n}$ and}
$$
H(\omega\|\shs\sigma)\leq\sum_{s=1}^{n-1}H(\omega_{A_s}).
$$
\end{lemma}\smallskip

\emph{Proof.} Let $\omega=|\Omega\rangle\langle\Omega|$. The Schmidt decomposition w.r.t. the subsystems $A_1$ and $A_2...A_n$ implies
$$
|\Omega\rangle=\sum_{i_1}\sqrt{p^1_{i_1}}|\varphi^1_{i_1}\rangle\otimes |\psi^1_{i_1}\rangle,
$$
where $\{\varphi^1_{i_1}\}$ and $\{\psi^1_{i_1}\}$ are orthogonal sets of unit vectors in $\H_{A_1}$ and $\H_{A_2...A_n}$ correspondingly and $\{p^1_{i_1}\}$ is a probability distribution.
By using the Schmidt decomposition of any vector of the set $\{\psi^1_{i_1}\}$  w.r.t. the subsystems $A_2$ and $A_3...A_n$  we obtain
$$
|\Omega\rangle=\sum_{i_1,i_2}\sqrt{p^1_{i_1}p^2_{i_1i_2}}|\varphi^1_{i_1}\rangle\otimes|\varphi^2_{i_1i_2}\rangle\otimes |\psi^2_{i_1i_2}\rangle,
$$
where $\{\varphi^2_{i_1i_2}\}_{i_2}$ and $\{\psi^2_{i_1i_2}\}_{i_2}$ are orthogonal sets of unit vectors in $\H_{A_2}$ and $\H_{A_3...A_n}$ correspondingly
and $\{p^2_{i_1i_2}\}_{i_2}$ is a probability distribution for any given $i_1$.

By repeating this process we get
\begin{equation*}
|\Omega\rangle=\sum_{i_1,i_2,...,i_{k-1}}\sqrt{p^1_{i_1}p^2_{i_1i_2}...p^{k-1}_{i_1i_2...i_{k-1}}}|\varphi^1_{i_1}\rangle\otimes...\otimes|\varphi^{k-1}_{i_1i_2...i_{k-1}}\rangle\otimes |\psi^{k-1}_{i_1i_2...i_{k-1}}\rangle,
\end{equation*}
for any $k\leq n$, where $\{p^{s}_{i_1i_2...i_{s}}\}_{i_{s}}$ and $\{\varphi^s_{i_1i_2...i_{s}}\}_{i_{s}}$, $s<k$, are, respectively,
a probability distribution and an orthogonal set of unit  vectors in $\H_{A_s}$ for any $i_1,i_2,...,i_{s-1}$ and
$\{\psi^{k-1}_{i_1i_2...i_{k-1}}\}_{i_{k-1}}$ is an  orthogonal set of unit vectors in $\H_{A_k...A_n}$ for any $i_1,i_2,...,i_{k-1}$.\smallskip

Consider the countably decomposable separable state
\begin{equation}\label{sigma-def}
\sigma=\sum_{i_1,i_2,...,i_{n-1}}p^1_{i_1}p^2_{i_1i_2}...p^{n-1}_{i_1i_2...i_{n-1}}\rho^1_{i_1}  \otimes...\otimes\rho^{n-1}_{i_1i_2...i_{n-1}}\otimes \varrho^{n-1}_{i_1i_2...i_{n-1}}
\end{equation}
in $\S(\H_{A_1...A_n})$, where $\rho^s_{i_1i_2...i_{s}}=|\varphi^{s}_{i_1i_2...i_{s}}\rangle\langle\varphi^{s}_{i_1i_2...i_{s}}|$, $s<n$, is a state in $\S(\H_{A_s})$ for any $i_1,i_2,...,i_{s}$
and $\varrho^{n-1}_{i_1i_2...i_{n-1}}=|\psi^{n-1}_{i_1i_2...i_{n-1}}\rangle\langle\psi^{n-1}_{i_1i_2...i_{n-1}}|$ is a state in $\S(\H_{A_n})$ for any $i_1,i_2,...,i_{n-1}$.

By noting that all the summands in (\ref{sigma-def}) are mutually orthogonal pure states one can show that the state $\sigma$ has the required properties.  $\square$

\bigskip

I am grateful to A.S.Holevo and G.G.Amosov for the discussion that motivated this research. I am also grateful to S.N.Filippov for the useful reference.

\end{document}